\documentclass[11pt]{article}
\usepackage[T1]{fontenc}
\usepackage{array}
\usepackage{amsmath}
\usepackage{amssymb}
\usepackage{graphicx}
\usepackage{authblk}
\usepackage{algorithm}
\usepackage{algorithmic}
\usepackage[authoryear]{natbib}
\usepackage[unicode=true,
 bookmarks=false,
 breaklinks=false,pdfborder={0 0 1},backref=section,colorlinks=false]
 {hyperref}
\newlength{\trianglewidth}
\newlength{\pluswidth}
\newlength\myindent
\newcommand\bindent{%
  \begingroup
  \setlength{\itemindent}{\myindent}
}
\newcommand\eindent{\endgroup}
\makeatletter

\providecommand{\tabularnewline}{\\}

\@ifundefined{date}{}{\date{}}

\usepackage{latexsym}
\usepackage{multirow}\usepackage{bm}
\usepackage{url}

\usepackage{enumerate}

\usepackage{colortbl}
\usepackage{subcaption}

\usepackage{dcolumn}
\newcolumntype{.}{D{.}{.}{-1}}
\newcolumntype{d}[1]{D{.}{.}{#1}}
\usepackage{theorem}
\newtheorem{assumption}{Assumption}\newtheorem{theorem}{Theorem}\newtheorem{lemma}{Lemma}

\usepackage{rotating}

\usepackage{arydshln}

\usepackage[compact]{titlesec}

\allowdisplaybreaks

\newcommand{\spacingset}[1]{\renewcommand{\baselinestretch}%
{#1}\small\normalsize}

\newcommand{\pr}{P}
\newcommand{\var}{\textnormal{var}}

\newcommand{\E}{\mathbb{E}}

\newcommand{\eff}{\mathrm{eff}}

\makeatother

\begin{document}
\title{Doubly Protected Estimation for Survival Outcomes Utilizing External Controls for Randomized Clinical Trials}

\author[1, 2]{Chenyin Gao}
\author[1]{Shu Yang\thanks{Corresponding author: syang24@ncsu.edu}}
\author[3]{Mingyang Shan}
\author[3]{Wenyu Wendy Ye}
\author[3]{Ilya Lipkovich}
\author[3]{Douglas Faries}
\affil[1]{North Carolina State University, Department of Statistics, Raleigh, NC, USA}
\affil[2]{Harvard University, Department of Biostatistics, Boston, MA, USA}
\affil[3]{Eli Lilly \& Company, Indianapolis, IN, USA}

\maketitle
\spacingset{1.5} 
\begin{abstract}
Censored survival data are common in clinical trials, but small control groups can pose challenges, particularly in rare diseases or where balanced randomization is impractical. Recent approaches leverage external controls from historical studies or real-world data to strengthen treatment evaluation for survival outcomes. However, using external controls directly may introduce biases due to data heterogeneity. We propose a doubly protected estimator for the treatment-specific restricted mean survival time difference that is more efficient than trial-only estimators and mitigates biases from external data. Our method adjusts for covariate shifts via doubly robust estimation and addresses outcome drift using the DR-Learner for selective borrowing. The approach can incorporate machine learning to approximate survival curves and detect outcome drifts without strict parametric assumptions, borrowing only comparable external controls. Extensive simulation studies and a real-data application evaluating the efficacy of Galcanezumab in mitigating migraine headaches have been conducted to illustrate the effectiveness of our proposed framework.
\end{abstract}

\noindent%
\vfill

\newpage

\section{Introduction}

Understanding the risk of disease or death and how these risks evolve over time is critical for assisting clinicians in treatment assignment and disease diagnosis. In clinical trials or biomedical studies, evaluating the effectiveness of drugs often suffers from limited sample sizes due to low disease prevalence or restrictive inclusion/exclusion criteria. This issue is exacerbated in survival analysis, as time-to-event or survival endpoints may not always be observed. As a complement to clinical trials, external controls offer a promising avenue to improve statistical inference when recruiting more patients is challenging. However, external controls can differ from clinical trials in many aspects due to differences in the underlying data acquisition and generation mechanisms. Concerns regarding the plausibility of these assumptions have limited their broader deployment. Guidance documents from regulatory agencies, including the recent Food and Drug Administration (FDA) draft guidance on Considerations for the Design and Conduct of Externally Controlled Trials for Drug and Biological Products, note several potential issues with the use of external controls, including selection bias, lack of concurrency, differences in the definitions of covariates, treatments, or outcomes, and unmeasured confounding \citep{FDA2001, FDA2019, FDA2023}. Each of these concerns can result in biased treatment effect estimates if external controls are integrated with the trial without further scrutiny.

\section{Related Work}\label{sec:related_work}


\paragraph{Data Integration with Non-survival Outcomes}
To reliably leverage external data, it is crucial to address these potential issues with the use of external data. One primary concern is the distributional heterogeneity of the baseline disease characteristics between the two studies, leading to the issue of covariate shifts. Likelihood-based frameworks have been explored to mitigate covariate shifts arising from external datasets \citep{chatterjee2016constrained,huang2016efficient}. Other 
propensity score weighting and matching methods have been proposed to construct new external data that have similar covariate distributions as the trial data \citep{li2018balancing}. However, these frameworks often rely on the invariance assumption, which posits that the conditional outcome distributions are identical for the trial and external controls. This assumption can be problematic, as the distributions of outcomes may also vary across studies given the baseline covariates, leading to the problem of outcome drift.

In recent years, various adaptive learning methods have been proposed to address the study-specific generating mechanisms for outcomes, aiming to ensure robust estimation even when external controls differ significantly from the trial. The considered analytic frameworks include adaptive information borrowing from diverse populations in linear regression \citep{li2022transfer, yang2023elastic, gao2023pretest}, generalized linear models \citep{tian2023transfer,li2023targeting}, and nonparametric classification \citep{cai2021transfer}. 

\paragraph{Data Integration with Survival Outcomes}
However, the adaptive learning in survival analysis remains limited due to challenges in incorporating external time-to-event outcomes. Unlike regression settings, where models dynamically borrow information for covariate effects, survival analyses need to address outcome drift in hazards as well. For instance, \citet{liu2014estimating} and \citet{huang2016efficient} proposed accommodating outcome heterogeneity by applying a constant factor to the cumulative hazard function. Additionally, \citet{chen2022propensity} and \citet{wang2020propensity} developed a propensity score-integrated Bayesian framework for the Kaplan–Meier (KM) estimator, which first selects external controls with similar hazard risks and then down-weights their impacts before incorporating them via the weighted KM estimator.

Nonetheless, these approaches overlook the time-varying nature of the outcome drift for survival analysis. To address this, \citet{chen2021combining} developed an adaptive empirical likelihood estimation that incorporates constraints from external summary-level information to account for time-varying baseline hazard differences. \citet{huang2023covariate} proposed a federated external control method to estimate hazard ratios in a federated weighted Cox model for time-to-event outcomes. Due to privacy and logistical concerns with data-sharing, these frameworks are designed to utilize external aggregated survival information, which can be restrictive and less efficient. 

Furthermore, it is essential to control the information sharing between covariate effects and hazard risks simultaneously. \citet{li2023accommodating} proposed a transfer learning framework that allows for comparable information borrowing in both covariate effects and baseline hazards through penalized likelihood under Cox models. However, this framework lacks flexibility in modeling time-varying covariate effects. \citet{bellot2019boosting} proposed learning the shared representation between two populations via the flexible nonparametric survival trees and correcting distribution mismatches with boosting, aiming to improve prediction performance without providing uncertainty quantification.


\paragraph{Our Contributions} Existing integrative methods are limited by the assumption of the Cox model, either on the cause-specific or subdistribution hazard scale, which requires to accurately model the survival curves. In recent years, semiparametric efficient and doubly robust estimators, which leverage the efficient influence function (EIF), including the methodology of solving the EIF-based estimation equation \citep{gao2023integrating, lee2024transporting} and the targeted maximum likelihood estimation \citep{rytgaard2022continuous, rytgaard2023estimation}, have gained great popularity to draw inferences about the treatment effects for survival outcomes. Therefore, there is a pressing need for developing a flexible and data-adaptive integrative framework that accounts for outcome drift in time-to-event outcomes, coupled with valid inferential methods, to enhance efficiency and reliability in survival model estimation.

In this paper, we develop a doubly protected estimation method for evaluating treatment effects for survival outcomes. To correct for covariate shifts, we utilize the density ratio of baseline covariates between two datasets in the construction of the doubly robust treatment estimator, motivated by the semi-parametric EIF. Since our framework is developed based on the EIF, it offers an advantage over other non-parametric methods in the construction of confidence intervals. 

Next, we recast this influence function into a selection-based integrative framework to address outcome drift, identifying a comparable subset of external data for borrowing. To adjust time-varying hazards for survival outcomes, we propose to detect this subset based on differences in the subject-level restricted mean survival time (RMST) with DR-Learner. By minimizing the bias-variance trade-off, our framework does not require stringent parametric assumptions on the survival curves and allows for the dynamic borrowing of external information with time-varying hazards. 

Finally, we establish the asymptotic properties of our proposed data-adaptive integrative estimator for survival outcomes with guaranteed consistency and efficiency improvement, even in the presence of external heterogeneity. Besides, we demonstrate the robustness and reasonable efficiency gains via extensive simulation studies and a real-data application. Our implementation codes will be made publicly available after the acceptance of this manuscript.

\section{Methodology}\label{sec:Problem-setting}

\subsection{Notation, Assumptions and Identifications}

Following the potential outcomes framework, let $T^{(a)}$ be the
potential survival time if a subject received the binary treatment
$A=a$. Let $S_{a}(t)$ and $\lambda_{a}(t)$ be the corresponding
survival and hazard functions, defined as $S_{a}(t)=\pr(T^{(a)}\geq t)$
and $\lambda_{a}(t)=\lim_{h\rightarrow0}h^{-1}\pr(t\leq T^{(a)}\leq t+h)/\pr(T^{(a)}\geq t)$. Under the consistency assumption, the observed survival time $T$
is the realization of potential outcome under the actualized treatment,
i.e., $T=AT^{(1)}+(1-A)T^{(0)}$. In the presence of censoring, the
survival time $T$ is not always observable. Instead, we observe $Y=\min(T,C)$
and $\Delta=\mathbf{1}(T<C)$, where $C$ is the censoring time, and
$\mathbf{1}(\cdot)$ is an indicator function. Let $M_{a}^{C}(dt\mid X,R=r)=dN_{a}^{C}(t)-\mathbf{1}(Y\geq t)\lambda_{a}^{C}(t\mid X,R=r)$
be a martingale with $dN_{a}^{C}(t)=\mathbf{1}(Y=t,\Delta=0,A=a)$, and $\lambda_{a}^{C}(t\mid X,R=r)=\lim_{h\rightarrow0}h^{-1}P(t\leq C\leq t+h\mid X,A=a,R=r)/P(C\geq t\mid X,A=a,R=r)$. 

Suppose we have two data sets: the trial data and the external controls.
Let $X$ be the baseline covariates and $R$ be the indicator of the data source, where $R=1$ if the
subject is from the trial data and $R=0$ for the external controls.
For the trial data, we observe $\mathcal{R}=\{V_{i}=(Y_{i},\Delta_{i},A_{i},X_{i},R_{i}=1)\}_{i=1}^{N_{t}}$;
for the external controls, only $\mathcal{E}=\{V_{i}=(Y_{i},\Delta_{i},A_{i}=0,X_{i},R_{i}=0)\}_{i=N_{t} +1}^{N_{t} + N_{e}}$
are observed since no treatment is assigned. Denote the true distribution
for $V_{i}$ by $\mathbb{P}$, and the empirical measure by $\mathbb{P}_{N}$
as $\mathbb{P}_{N}(f)=\sum_{i}f(V_{i})/N$, where $N=N_{t}+N_{e}$.

Let $\pi_R(X) = P(R=1\mid X)$, $q_{R}(X)=\pi_R(X)/\{1-\pi_R(X)\}$ be the density ratio of the baseline covariates, $\pi_{A}(X)=P(A=1\mid X,R=1)$ be the propensity score for the treatment, and $S_{a}(t\mid R=1)=\pr(T^{(a)}\geq t\mid R=1)$ be the treatment-specific
survival curves for the trial population. The parameter of our interest
$\theta_{\tau}$ is the average treatment effect measured by the restricted
mean survival time (RMST) difference up to $\tau$, defined by $\theta_{\tau}=\int_{0}^{\tau}\{S_{1}(t\mid R=1)-S_{0}(t\mid R=1)\}dt$.
To identify the difference in RMST, the following assumptions are
sufficient.

\begin{assumption}[Internal validity for the trial data]\label{assum:trial}
(i) $T^{(a)}\perp A\mid X,R=1$ for $a=0,1$; and (ii) $0<\pi_{A}(X), \pi_R(X)<1$ in the
support of $X$.
\end{assumption}

\begin{assumption}[Informative censoring]\label{assum:censoring_noninformative}
$T^{(a)}\perp C\mid X,A=a,R=r$ for $a=0,1$ and $r=0, 1$.
\end{assumption}

\begin{assumption}[Comparability for the external data]\label{assum:exchangeability_mean}
$S_{0}(t\mid X)=S_{0}(t\mid X,R=0)=S_{0}(t\mid X,R=1)$ for any $t<\tau$
in the support of $X$, where $S_{a}(t\mid X,R)=\pr(T^{(a)}\geq t\mid X,R)$; 
\end{assumption}

Assumption \ref{assum:trial} holds by the design of trial and is useful to detect the external
heterogeneity. Assumption
\ref{assum:censoring_noninformative} is a common censoring at random
assumption for survival analysis, which is a special case of the coarsening
at random \citep{tsiatis2006semiparametric}. Assumption \ref{assum:exchangeability_mean}
states the external data is comparable to the trial data if there is a rich set of covariates capturing all the
outcome
predictors that are correlated with the data source indicator $R$. 

However, Assumption \ref{assum:exchangeability_mean} is prone to violations in practice due to many bias-generating concerns, such as unmeasured confounding, lack of concurrency, and outcome validity. Our proposed framework is two-fold: 1) Under Assumption \ref{assum:exchangeability_mean} where covariate shift can be present, we develop a semi-parametric efficient integrative estimator for the treatment effects evaluation using the combined data sets (Section \ref{subsec:full borrowing}); 2) Considering the potential violation of Assumption \ref{assum:exchangeability_mean} where the outcome drift is allowed, we adapt the efficient estimation into a selective integrative procedure that first detects the biases and only retains a subset of comparable external data for integration (Section \ref{subsec:selective borrowing}).

\subsection{Efficient Integrative Estimation Assuming Population Homogeneity}\label{subsec:full borrowing}
Under Assumptions \ref{assum:trial} to \ref{assum:exchangeability_mean}, the average treatment effects $\theta_{\tau}$, or $S_{a}(t\mid R=1)$
sufficiently, is identified based on the observed data. The following
theorem provides the identification formulas.

\begin{theorem}\label{thm:identification0}

Under Assumptions \ref{assum:trial} to \ref{assum:exchangeability_mean},
the following identification formulas hold for the treatment-specific
survival curves $S_{a}(t\mid R=1)$.
\begin{enumerate}
\item [(a)] Based on the trial data: 
\begin{align*}
&S_{a}(t\mid R=1)  =\frac{1}{\pr(R=1)}\E\left\{ RS_{a}(t\mid X,R=1)\right\} =\frac{1}{\pr(R=1)}\E\left\{ \frac{R\mathbf{1}(A=a)\Delta\mathbf{1}(Y>t)}{P(A=a\mid X)\pi^{C}_1(Y, X)}\right\},
\end{align*}
where $\pi^{C}_1(t, X)=P(C\geq t\mid X, R=1)$ is the censoring probability for the trial. 
\item [(b)] Based on the external data:
\begin{align*}
& S_{0}(t\mid R=1) =\frac{1}{\pr(R=1)}\E\left\{ \frac{(1-R)q_{R}(X)\Delta\mathbf{1}(Y>t)}{\pi^{C}_0(Y, X)}\right\},
\end{align*}
where $\pi^{C}_0(t, X)=P(C\geq t\mid X, R=0)$ is the censoring probability for the external controls. 
\end{enumerate}
\end{theorem}

Theorem \ref{thm:identification0} provides the identification formulas
for $S_{a}(t\mid R=1)$, which is sufficient to identify the average
treatment effect $\theta_{\tau}$ among the trial population. In particular,
Theorem \ref{thm:identification0}(a) identifies $S_{a}(t\mid R=1)$
based on the outcome imputation or the inverse probability censoring
weighting with the trial data only; Theorem \ref{thm:identification0}(b) uses the external controls to identify $S_{0}(t\mid R=1)$. As
the covariate distribution of the external controls may not be representative
of the trial population, the identification formula relies on the
density ratio $q_{R}(X)$ to obtain the survival curves marginalized
over the trial population. A detailed proof of Theorem \ref{thm:identification0} is provided in Appendix \ref{subsec:proof_identification0}. 

However, the identification formulas provided in Theorem \ref{thm:identification0}
could motivate infinitely many estimators for $\theta_{\tau}$ under
Assumptions \ref{assum:trial} to \ref{assum:exchangeability_mean}.
To construct a more principled estimator, we derive the efficient
influence function (EIF) of $\theta_{\tau}$ in Theorem \ref{thm:EIF0}
based on the semiparametric theory \citep{tsiatis2006semiparametric}. The EIF, also known as the canonical gradient \citep{van2011targeted}, is a fundamental tool to achieve local semiparametric efficiency for estimation. 

\begin{theorem} \label{thm:EIF0}
Under Assumptions \ref{assum:trial} to \ref{assum:exchangeability_mean}, 
\begin{enumerate}
    \item [(a)] the EIF for $S_{1}(t\mid R=1)$ is $\psi_{S_1, \eff}(t,V)=\phi_{S_{1},\eff}(t,V) - RS_1(t\mid R=1)/\pr(R=1)$, where
\begin{align*}
&\phi_{S_{1},\eff}(t,V) =\frac{RA\mathbf{1}(Y>t)}{\pr(R=1)\pi_{A}(X)\pi^C_1(t, X)} +\int_{0}^{t}\frac{RA\cdot dM_{1}^{C}(r\mid X)}{\pr(R=1)\pi_{A}(X)\pi^C_1(r, X)}\frac{S_{1}(t\mid X,R=1)}{S_{1}(r\mid X,R=1)}\\
 &+\frac{RS_{1}(t\mid X,R=1)}{\pr(R=1)}\left\{ 1-\frac{A}{\pi_{A}(X)}\right\}.
\end{align*}
\item [(b)] the EIF for $S_{0}(t\mid R=1)$ is $\psi_{S_0, \eff}(t,V)=\phi_{S_{0},\eff}(t,V) - RS_0(t\mid R=1)/\pr(R=1)$, where 
\begin{align*}
& \phi_{S_{0},\eff}(t,V) =\frac{R(1-A)}{\pr(R=1)}\frac{q_{R}(X)\mathbf{1}(Y>t)}{\pi^C_1(t, X)D(t,X)}\\
 & +\int_{0}^{t}\frac{R(1-A)}{\pr(R=1)}\frac{q_{R}(X)dM_{0}^{C}(r\mid X,R=1)}{\pi^C_1(r, X)D(t,X)}\times\frac{S_{0}(t\mid X,R=1)}{S_{0}(r\mid X,R=1)}\\
 & +\int_{0}^{t}\frac{(1-R)}{\pr(R=1)}\frac{q_{R}(X)r(t,X)dM_{0}^{C}(r\mid X,R=0)}{\pi^C_0(r, X)D(t,X)}\times \frac{S_{0}(t\mid X,R=0)}{S_{0}(r\mid X,R=0)}\\
&+\frac{(1-R)}{\pr(R=1)}\frac{q_{R}(X)r(t,X)\mathbf{1}(Y>t)}{\pi^C_0(t, X)D(t,X)} \\
& +\frac{S_{0}(t\mid X,R=1)}{\pr(R=1)}\left\{ \frac{Rq_{R}(X)\{A-\pi_{A}(X)\}}{D(t,X)} +\frac{r(t,X)\{R-(1-R)q_{R}(X)\}}{D(t,X)}\right\},
\end{align*}
$D(t,X)=r(t,X)+\{1-\pi_{A}(X)\}q_{R}(X)$, $r(t,X)=V_{R1,A0}/V_{R0}$, $V_{R1,A0}=  \var\left\{ \mathbf{1}(T>t)\mid R=1,A=0\right\}$, and $V_{R0}=\var\left\{ \mathbf{1}(T>t)\mid R=0\right\}$.
\item [(c)] the EIF for $\theta_{\tau}$ is:
$$
\psi_{\theta_{\tau}, \eff}(V) = \int_0^\tau \{\psi_{S_{1},\eff}(t,V) - \psi_{S_{0},\eff}(t,V)\}dt.
$$
\end{enumerate}
\end{theorem}
Theorem \ref{thm:EIF0}(a) shows the EIF for $S_{1}(t\mid R=1)$ with the trial data, which is well-studied as the observed-data EIF under monotone coarsening in \cite{tsiatis2006semiparametric}; Theorem \ref{thm:EIF0}(b) shows the EIF for $S_{0}(t\mid R=1)$, which
is an extension of \citet{gao2023integrating} with additional integral terms contributed by the censoring scores of concurrent controls and external data; Theorem
\ref{thm:EIF0}(c) suggests that the EIF for $\theta_{\tau}$ is an  integral of the EIFs for $S_a(t\mid R=1)$ as the average treatment effect $\theta_{\tau}$ is an integral
function of the treatment-specific survival curves $S_{a}(t\mid R=1)$. Furthermore, the proposed integrative framework can be generalized to any estimand that is a function of the survival function $S_a(t)$ (e.g., the mean or median of the survival time), with only a trivial algebraic extension, since the EIFs are derived for $S_a(t)$.

We now give some intuitions
behind the EIF $\psi_{S_{0},\eff}(t,V)$ for the combined data sets. First,
we derive the full-data EIF $\psi_{S_{0},\eff}^{F}(t,V)$ under Assumption \ref{assum:exchangeability_mean}, which involves a part being the weighted-averaged of the EIFs for the trial and external data with weights being inversely proportionately
to the variances for the EIFs of the trial and external data, respectively. Next, we project
the full-data EIF $\psi_{S_{0},\eff}^{F}(t,V)$ to the coarsened data
space induced by the censoring, and gives the observed-data EIF in Theorem \ref{thm:EIF0}(b); see Theorem 10.4, \citet{tsiatis2006semiparametric}. A detailed proof of Theorem \ref{thm:EIF0} is provided in Appendix \ref{sec:proof_EIF0}. 

However, constructing an estimator based on the EIFs first requires approximating the unknown nuisance functions $\pi_{R}(X)$, $\pi_{A}(X)$, $S_{a}(t\mid X)$,
and $S^{C}(t\mid X,R)$. By replacing the true distribution $\mathbb{P}$
with its estimated counterparts and solving the empirical expectation
$\mathbb{P}_{N}\{\widehat{\psi}_{\theta_{\tau},\eff}(V)\}$, we have
\begin{align*}
\widehat{\theta}_{\tau}^{\text{acw}} & =N^{-1}\sum_{i\in\mathcal{R}\cup\mathcal{E}} \int_{0}^{\tau} \left\{\widehat{\phi}_{S_{1},\eff}(t,V_{i})dt - \widehat{\phi}_{S_{0},\eff}(t,V_{i})\right\}dt.
\end{align*}
Under the conditions in Theorem \ref{thm:theta_dr}, our proposed
integrative estimator $\widehat{\theta}_{\tau}^{\text{acw}}$ is rate doubly robust, asymptotically
normal, and locally efficient as established below.

\begin{theorem}\label{thm:theta_dr}
Denote
$\|f\|_{L_{2}}=\mathbb{P}\{f(V)^{2}\}^{1/2}$ where $\mathbb{P}$
is the true distribution. Let $\widehat{\pi}_{R}(X)$, $\widehat{\pi}_{A}(X)$,
$\widehat{S}_{a}(t\mid X,R=1)$, and $\widehat{\pi}^{C}_r(t, X)$
be general semi-parametric models for $\pi_{R}(X)$, $\pi_{A}(X)$,
$S_{a}(t\mid X,R=1)$, and $\pi^{C}_r(t, X)$, respectively. Suppose Assumptions \ref{assum:trial} to \ref{assum:exchangeability_mean} and the regularity conditions \ref{assump:rate} are satisfied, up to a multiplicative constant, the estimation error $\widehat{\theta}_{\tau}^{\text{acw}}-\theta_{r}=\mathrm{err}(\widehat{\theta}_{\tau}^{\text{acw}}, \theta_{r})$ is
bounded by
\begin{align*}
& \mathrm{err}(\widehat{\theta}_{\tau}^{\text{acw}}, \theta_{r}) =  \left\{ \|\widehat{S}_{0}(t\mid X,R=1)-S_{0}(t\mid X,R=1)\|_{L_{2}}+\|\widehat{S}_{1}(t\mid X,R=1)-S_{1}(t\mid X,R=1)\|_{L_{2}}\right\} \\
 & \times\left\{ \|\widehat{\pi}_{A}(X)-\pi_{A}(X)\|_{L_{2}}+\|\widehat{\pi}_{R}(X)-\pi_{R}(X)\|_{L_{2}}  +\max_{r<t}\|\widehat{\lambda}_{0}^{C}(r\mid X,R)-\lambda_{0}^{C}(r\mid X,R)\|_{L_{2}}\right\}.
\end{align*}
Thus, we have $N^{1/2}(\widehat{\theta}_{\tau}^{\text{acw}}-\theta_{\tau})\overset{d}{\rightarrow}N(0,\mathbb{V}_{\tau})$,
where $N=N_{t}+N_{e}$ and $\mathbb{V}_{\tau}=\E\{\psi_{\theta_{\tau},\eff}^{2}(V)\}$
is the semi-parametric lower bound for the variance.
\end{theorem}
Theorem \ref{thm:theta_dr} shows that the proposed estimator $\widehat{\theta}_{\tau}$ can incorporate
flexible nonparametric or machine learning methods for estimating the nuisances with the
required convergence rates, while maintaining the parametric-rate consistency. The proof of Theorem \ref{thm:theta_dr} is deferred to Appendix \ref{sec:prrof_dr}.

\subsection{Robust Selective Borrowing with DR-Learner for Outcome Drift Detection}\label{subsec:selective borrowing}

In practice, Assumption \ref{assum:exchangeability_mean} is often
violated, that is, $S_0(t\mid X,R=1)\neq S_0(t\mid X,R=0)$ for some $t\in[0,\tau]$,
and $\widehat{\theta}_{\tau}$ may be biased. Suppose
there exists a comparable subset $\mathcal{A}\subseteq\{1,\cdots,N_{e}\}$
of the external controls such that $S_0(t\mid X_{i},R=1)=S_0(t\mid X_{i},R=0)$
for any time $t$ and subject $i\in\mathcal{A}$, and we aim to selectively
borrow this comparable subset.
However, the subset $\mathcal{A}$ is often unknown a priori. In this section, we propose a robust selective borrowing framework to incorporate comparable external controls in estimating the average treatment effect. We introduce a vector of bias parameter $b_{0}=(b_{1,0},\cdots,b_{N_{e},0})$,
where $b_{i,0}=\int_{0}^{\tau}S_{0}(t\mid X_{i},R_{i}=1)dt-\int_{0}^{\tau}S_{0}(t\mid X_{i},R_{i}=0)dt$.
To prevent bias in $\widehat{\theta}_{\tau}$, our goal is to identify the zero-valued subset of the bias parameter and leverage only this subset
for the integrative estimation. 

One simple estimator to detect
the bias is the ``plug-in'' estimation, defined as $\widehat{b}_{i}=\widehat{b}(V_{i})=\int_{0}^{\tau}\widehat{S}_{0}(t\mid X_{i},R_{i}=1)dt-\int_{0}^{\tau}\widehat{S}_{0}(t\mid X_{i},R_{i}=0)dt$.
However, the ``plug-in'' estimator might have large finite-sample biases if one uses flexible models for the conditional survival functions. Heuristically, bias detection is equivalent to estimating the conditional differences in expected control means over two datasets, which parallels the conditional average treatment estimation in the causal inference literature when the study source indicator is perceived as the treatment indicator. To more accurately approximate the biases, the DR-Learner approach \citep{kennedy2020optimal,kallus2023robust} is utilized to construct the initial pseudo-outcome $\xi_{i}$ for the bias $b_{i,0}$ by $\xi_{i}=\xi(V_{i})=\int_{0}^{\tau}\kappa_{0}(t,V_{i}\mid R=1)dt-\int_{0}^{\tau}\kappa_{0}(t,V_{i}\mid R=0)dt$, where
\begin{align*}
&\kappa_{0}(t,V_{i}\mid R=1)  =S_{0}(t\mid X_{i},R=1) +\int_{0}^{t}\frac{R_{i}(1-A_{i})dM_{0}^{C}(r\mid X_{i},R=1)}{\pi_{R}(X_{i})\{1-\pi_{A}(X_{i})\}\pi^{C}_1(r, X)}\frac{S_{0}(t\mid X_{i},R=1)}{S_{0}(r\mid X_{i},R=1)}\\
 &+\frac{R_{i}(1-A_{i})}{\pi_{R}(X_{i})\{1-\pi_{A}(X_{i})\}}\times \left\{ \frac{\mathbf{1}(Y_{i}>t)}{\pi^{C}_1(t, X_{i})}-S_{0}(t\mid X_{i},R=1)\right\},
\end{align*}
and 
\begin{align*}
&\kappa_{0}(t,V_{i}\mid R=0)  =S_{0}(t\mid X_{i},R=0) +\int_{0}^{t}\frac{(1-R_{i})dM_{0}^{C}(r\mid X_{i},R=0)}{\{1-\pi_{R}(X_{i})\}\pi^{C}_0(t, X_{i})}\frac{S_{0}(t\mid X_{i},R=0)}{S_{0}(r\mid X_{i},R=0)}\\
 &+\frac{(1-R_{i})}{1-\pi_{R}(X_{i})}\left\{ \frac{\mathbf{1}(Y_{i}>t)}{\pi^{C}_0(t, X_{i})}-S_{0}(t\mid X_{i},R=0)\right\}.
\end{align*}
Lemma \ref{lemma:pseudo-outcomes} provides bounds for estimation 
 error for the pseudo-outcomes,
which is the key step for bias detection.

\begin{lemma}\label{lemma:pseudo-outcomes}

Let $\widehat{\xi}(V)$ be the pseudo-outcome with unknown nuisance functions replaced by their estimated counterparts and $\xi^{*}(V)$ be its probability limit,
the conditional expectation $\|\E\{\xi^{*}(V)-b_{0}\mid X\}\|_{L_{2}}$
is bounded by 
\begin{align*}
 & \sum_{r=0}^{1}\|\widehat{\pi}_{R}(X)-\pi_{R}(X)\|_{L_{2}}\times \|\widehat{S}_{0}(t\mid X,R=r)-S_{0}(t\mid X,R=r)\|_{L_{2}}\\
 & +\sum_{r=0}^{1}\|\widehat{\pi}^{C}_r(t,X)-\pi^{C}_r(t,X)\|_{L_{2}}\times \|\widehat{S}_{0}(t\mid X,R=r)-S_{0}(t\mid X,R=r)\|_{L_{2}}\\
 & +\|\widehat{\pi}_{A}(X)-\pi_{A}(X)\|_{L_{2}}\times \|\widehat{S}_{0}(t\mid X,R=1)-S_{0}(t\mid X,R=1)\|_{L_{2}},
\end{align*}
up to a multiplicative constant.
\end{lemma} 
The results in Lemma \ref{lemma:pseudo-outcomes} show that nuisance errors will have smaller impacts on the pseudo-outcomes $\widehat{\xi}$ as its estimation error is bounded by a quadratic form of the nuisance errors. The proof is provided in Appendix \ref{sec:pseudo}.

We now present our adaptive integrative framework in Algorithm \ref{alg:example}. The cutoff value $\tau$ for computing the RMST is crucial in practice, as the distribution of the tail beyond this point is neglected. Typically, the event rates at this cutoff value should exceed 10\% to ensure sufficient data for model development.

\begin{algorithm}[!htbp]
   \caption{Doubly Protected Adaptive Integrative Analysis for Survival Outcomes}
   \label{alg:example}
\begin{algorithmic}
   \STATE {\bfseries Input:} trial data 
   $ \mathcal{R}=\{V_{i}=(Y_{i},\Delta_{i},A_{i},X_{i},R_{i}=1)\}_{i=1}^{N_{t}}$, the external controls $\mathcal{E}=\{V_{i}=(Y_{i},\Delta_{i},A_{i}=0,X_{i},R_{i}=0)\}_{i=N_{t} +1}^{N_{t} + N_{e}}$, penalty function $p(\cdot)$, tuning parameter $\lambda_N$, and the cutoff value $\tau$.
   \STATE \COMMENT{Preparation}
\bindent
   \STATE Randomly split the data $\mathcal{R}\cup \mathcal{E}$ into two folds $\mathcal{I}_1$ and $\mathcal{I}_2$.
\eindent
\STATE \COMMENT{Step 1}
\bindent
    \STATE Fit the conditional survival curves $\widehat{S}_a(t\mid X, R=r)$ and $\widehat{S}^C(t\mid X, R=r)$ on $\mathcal{I}_1$ for $r=0, 1$.  
    \STATE Fit the propensity scores $\widehat{\pi}_R(X)$ and $\widehat{\pi}_A(X)$ on $\mathcal{I}_1$.
    \STATE Compute the pseudo-outcomes $\widehat{\xi}_i$ by $\kappa_{0}(t,V_{i}\mid R=1)$ and $\kappa_{0}(t,V_{i}\mid R=0)$ on $\mathcal{I}_2$.
\eindent
   \STATE \COMMENT{Step 2}
\bindent
    \STATE Refine the pseudo-outcomes by
    $\tilde{b}_{i}=\arg\min_{b_{i}}(\widehat{\xi}_{i}-b_{i})^{2}+\lambda_{N}p(|b_{i}|).
    $
    \STATE Obtain the comparable set $\tilde{\mathcal{A}}=\{i:\tilde{b}_{i}=0\}$. 
\eindent
   \STATE \COMMENT{Step 3}
\bindent
   \STATE Output the adaptive integrative estimator $\widehat{\theta}_{\tau}^{\text{adapt}}$.
\eindent
\end{algorithmic}
\end{algorithm}

In Algorithm \ref{alg:example}, Step 1 is a typical strategy to use machine learning techniques for estimating the nuisance functions; The penalty term in Step 2 is chosen to guarantee the selection consistency
such that $\pr(\tilde{\mathcal{A}}=\mathcal{A})\rightarrow1$, where
$\tilde{\mathcal{A}}=\{i:\tilde{b}_{i}=0\}$ is the estimated comparable
set. For example, the penalty term $p(\cdot)$ can be the adaptive
lasso \citep{zou2006adaptive}, the smoothly clipped absolute deviation (SCAD) \citep{fan2001variable}, and the minimax concave penalty (MCP) \citep{zhang2010nearly}; Finally, Step 3 outputs the adaptive integrative estimator $\widehat{\theta}_{\tau}^{\text{adapt}}$ with the comparable set $\tilde{\mathcal{A}}$ by
\begin{align*}
\widehat{\theta}_{\tau}^{\text{adapt}} &=N^{-1} \sum_{i\in\mathcal{R}\cup\mathcal{E}} \int_{0}^{\tau}\widehat{\phi}_{S_{1},\eff}(t,V_{i})dt-N^{-1} \sum_{i\in\mathcal{R}\cup\mathcal{E}} \int_{0}^{\tau}\widehat{\phi}_{S_{0},\eff}^{\tilde{\mathcal{A}}}(t,V_{i})dt,
\end{align*}
where the difference between $\phi_{S_{0},\eff}^{\mathcal{A}}(t,V)$ in Step 3 and $\phi_{S_{0},\eff}(t,V)$ in Theorem \ref{thm:EIF0} lies in the focus on leveraging the comparable set $\mathcal{A}$ instead of the whole external controls $\mathcal{E}$ for the integrative analysis.
\begin{theorem}\label{thm:eff_gain} Let $\mathbb{V}_{\tau}^{\text{aipw}}$ be the variance of the trial-only estimator. Under the same conditions in
Theorem \ref{thm:theta_dr}, we have $N^{1/2}(\widehat{\theta}_{\tau}^{\text{adapt}}-\theta_{\tau})\overset{d}{\rightarrow}N(0,\mathbb{V}_{\tau}^{\text{adapt}})$,
where 
\begin{align*}
\mathbb{V}_{\tau}^{\text{adapt}}-\mathbb{V}_{\tau}^{\text{aipw}}&=\frac{\pi_{R}(X)r(t,X)\pr(b_{0}=0\mid X,R=0)}{\pr(R=1)^{2}D_{b_{0}}(t,X)\{1-\pi_{A}(X)\}}\times \frac{D_{b_{0}}^{*}(t,X)}{D_{b_{0}}(t,X)}\frac{r(t,X)}{r^{*}(t,X)}V_{R1,A0},
\end{align*}
where $D_{b_{0}}^{*}(t,X)=r^{*}(t,X)\pr(b_{0}=0\mid X,R=0)+\{1-\pi_{A}(X)\}q_{R}(X)$ and $r^{*}(t,X)=V_{R1,A0}^{*}/V_{R0}^*$; $V_{R_1, A_0}$ are defined in Theorem \ref{thm:EIF0}, and the modified variance terms $V_{R1,A0}^{*}$ and $V_{R0}^{*}$ are defined in Appendix \ref{sec:proof_gain}.
\end{theorem}
Theorem \ref{thm:eff_gain} highlights the benefit of selective incorporating
external controls, where the asymptotic variance of $\widehat{\theta}_{\tau}^{\text{adapt}}$
is strictly smaller than the variance $\mathbb{V}_{\tau}^{\text{aipw}}$ of the trial-only estimator unless the external
study is in extremely poor quality (i.e., $r(t,X)=0$) or the comparable
external subset is empty (i.e., $\pr(b_{0}=0\mid X,R=0)=0$). A proof is provided in Appendix \ref{sec:proof_gain}.

\section{Simulation}\label{sec:Simulation}

In this section, we conduct several simulation studies to evaluate
the finite-sample performance of the proposed selective integrative
estimator. The sample size for the external controls are fixed at $N_{e}=500$,
and the parameter of our interest is the difference in RMST with $\tau=2$.
First, we generate $X=(X_{i,1},\cdots,X_{i,p})\sim\mathcal{N}(0,I_{p\times p})$
with $p=3$ for each subject $i$. Next, we generate the data source
indicator $R$ by Bernoulli sampling in Table \ref{tab:Summary-of-simulated}, where $\alpha_{R,0}$ is chosen such that the average of $R$ is around
$N_{t}/N$. For the trial data, where $R=1$, the
treatment $A$ is generated by Bernoulli sampling: 
\[
A\mid X,R=1\sim\text{Bernoulli}\left\{ \frac{\exp(\alpha_{A,0}+1_{p}^{\intercal}X)}{1+\exp(\alpha_{A,0}+1_{p}^{\intercal}X)}\right\} ,
\]
where $\alpha_{A,0}$ is chosen such that the average of $A$ is around
$N_{1}/N_{t}$, and $N_1$ is the size of treatment group. We consider the following conditional hazard function
$\lambda_{a}(t\mid X,R=1)=\exp(-.5a-1_{p}^{\intercal}X\cdot0.2)$, and $\lambda^{C}(t\mid X,R=1)=\exp(1_{p}^{\intercal}X\cdot0.1+\beta_{C})$ for the
trial. The time-to-event outcomes $T$ and the censoring times $C$ are generated
by inverting the survival functions induced by the hazard function
$\lambda_{a}(t\mid X)$ and $\lambda^{C}(t\mid X)$, respectively.
The parameter $\beta_{C}=1$ controls the expected censoring
time for the trial data where the censoring rates $P(C<T)$ is around
$40\%$; additional simulations where the data is subject to a different
intensities of censoring with various values of $\beta_{C}$ are available
in Appendix \ref{sec:additional}.

Next, to mimic the bias-generating concerns raised by the FDA, namely, selection bias, unmeasured confounding, lack of concurrency, and different covariate effects and time-varying baseline hazards, we consider five simulation settings for the external controls, as summarized in Table \ref{tab:Summary-of-simulated}. Under Setting One, the survival and censoring times for the external controls are generated using the exact same parameters as $\lambda_{a}(t \mid X)$ with $a$ fixed at 0 (i.e., no treatment). Under Setting Two, the hazard functions for both studies are confounded by an unobserved factor $U$ to maintain the same level of hazard variability across the two datasets. In particular, $U \sim \mathcal{N}(0,1)$ (zero mean) is included in the hazard function for the trial data, whereas $U + 1 \sim \mathcal{N}(1,1)$ (non-zero mean) is used for the external controls, which is expected to introduce greater outcome drift. Under Setting Three, the external controls are subject to inconcurrency bias, where $\delta_i$ takes values from $\{0, 5\}$ with equal probability $1/2$. Lack of concurrency could occur when the trial data and external control data are collected during different time periods or under varying healthcare settings. Under Setting Four, the external controls have different covariate effects, whereas under Setting Five, they exhibit different baseline time-varying hazards. Such discrepancies may arise if the two populations respond differently to the treatment (or placebo), even when they share the same baseline covariates.

\begin{table*}[!htbp]
\begin{center}
\caption{Summary of considered three simulation settings\label{tab:Summary-of-simulated}}
\vspace{0.15cm}
\begin{tabular}{p{.15\textwidth}p{.25\textwidth}p{.6\textwidth}}
\hline 
Bias & Setting & \multicolumn{1}{c}{Details}\tabularnewline
\hline 
Covariate shift &  \multirow{2}*{Selection bias} & 
$R\mid X\sim\text{Bernoulli}\left\{ \frac{\exp(\alpha_{R,0}+1_{p}^{\intercal}X)}{1+\exp(\alpha_{R,0}+1_{p}^{\intercal}X)}\right\}$,\\
&& $\lambda_{0}(t\mid X,R=0)=\exp(-1_{p}^{\intercal}X\cdot0.2)$\\
\hline
Outcome drift  & \multirow{2}*{Unmeasured confounder} & 
$R\mid X\sim\text{Bernoulli}\left\{ \frac{\exp(\alpha_{R,0}+1_{p}^{\intercal}X+U)}{1+\exp(\alpha_{R,0}+1_{p}^{\intercal}X+U)}\right\}$,\\
& & $\lambda_a(t\mid X,U,R)=\exp\{-0.5a-1_{p}^{\intercal}X\cdot0.2+3(U+\mathbf{1}(R=0)\}$\\
\\
& \multirow{2}*{Lack of concurrency} & $R\mid X\sim\text{Bernoulli}\left\{ \frac{\exp(\alpha_{R,0}+1_{p}^{\intercal}X)}{1+\exp(\alpha_{R,0}+1_{p}^{\intercal}X)}\right\}$, \\
&& $\lambda_{a}(t\mid X,R)=\exp\{-.5a-1_{p}^{\intercal}X\cdot0.2+3\delta\mathbf{1}(R=0)\}$\\
\\
& \multirow{2}*{Different covariate effect} & $\lambda_a(t\mid X, R=1) = \exp(-.5a-1_{p}^{\intercal} X \cdot 0.2)$\\
& & $\lambda_0(t\mid X, R=0) = \exp(-1_{p}^{\intercal} X \cdot 0.5)$\\
\\
& \multirow{1}*{Different baseline hazard} & $\lambda_a(t\mid X, R=1) = t\exp(-.5a-1_{p}^{\intercal} X \cdot 0.2)$ \\
& & $\lambda_0(t\mid X, R=0) = 2t\exp(-1_{p}^{\intercal} X \cdot 0.2)$ \\
\hline
\end{tabular}\end{center}
\end{table*}

In our evaluation, we compare the proposed selective borrowing estimator
$\widehat{\theta}_{\tau}^{\text{adapt}}$ with the trial-only estimator
$\widehat{\theta}_{\tau}^{\text{aipw}}$ \citep{tsiatis2006semiparametric}, the naive full borrowing
estimator $\widehat{\theta}_{\tau}^{\text{acw}}$, the propensity
score-integrated estimator $\widehat{\theta}^{\text{psrwe}}$ \citep{chen2020propensity},
and the transfer learning Cox regression $\widehat{\theta}_{\tau}^{\text{TransCox}}$
\citep{li2023accommodating}. The penalty term $p(\cdot)$ is chosen to be the adaptive lasso \citep{zou2006adaptive}. The conditional survival curves $S_a(t \mid X)$ and $S^C(t \mid X)$ are modeled by the Cox model, and the propensity scores $\pi_R(X)$ and $\pi_A(X)$ are modeled by the SuperLearner with the logistic regression and random forest as the base learners \citep{van2006targeted}.

\begin{figure}[!htbp]
\centering
\includegraphics[width=.6\linewidth]{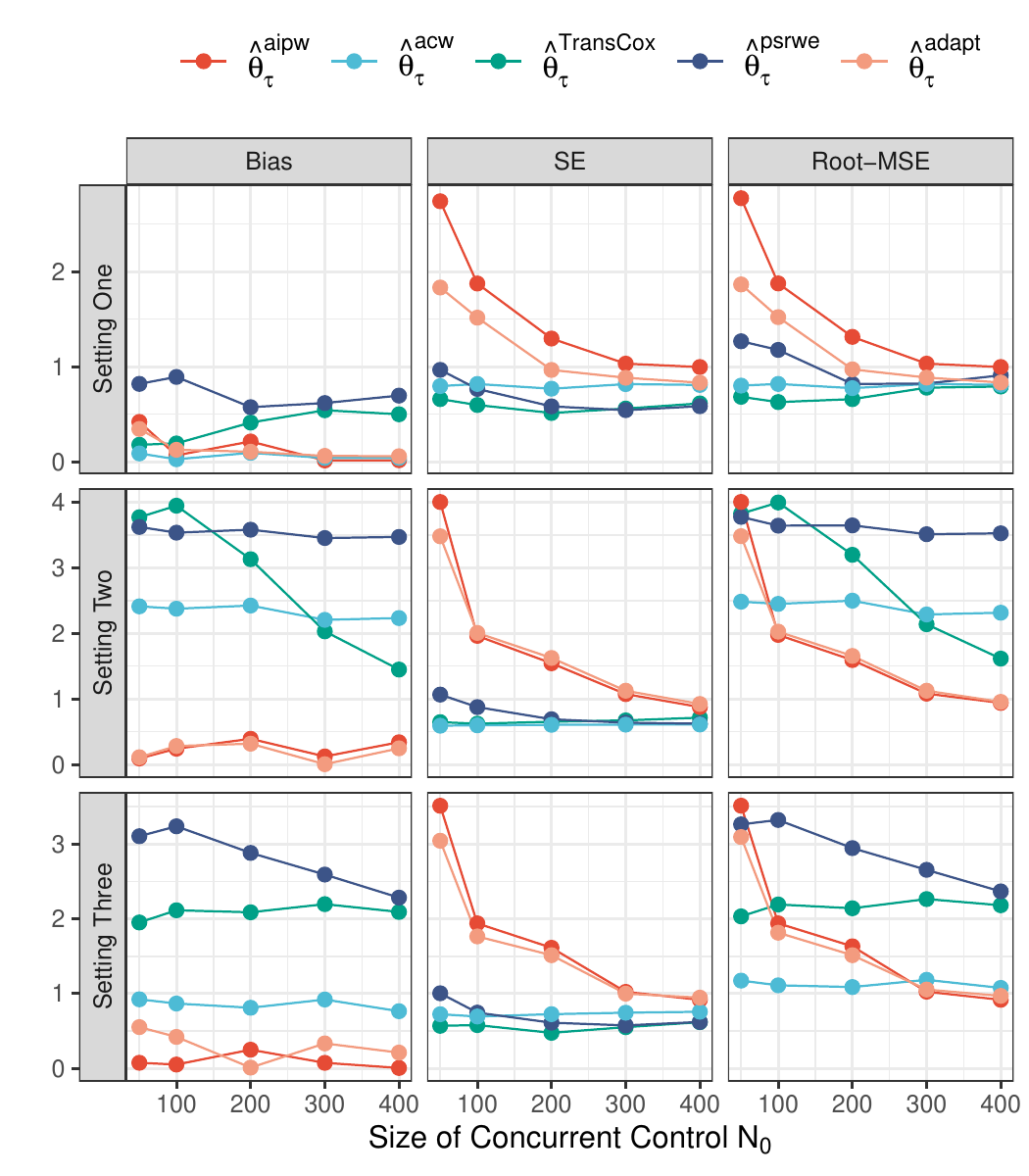}
\caption{\label{fig:sim-point} Point estimation results for RMST over $500$
Monte Carlo experiments.}
\end{figure}

Figure \ref{fig:sim-point} presents
the bias, standard error (SE), and the square root of the mean squared
error (Root-MSE) for each method across three simulation settings,
with the size of the concurrent control $N_{0}$ ranging from $50$
to $400$. As expected, the trial-only benchmark estimator $\widehat{\theta}_{\tau}^{\text{aipw}}$
exhibits small biases across these three settings by design. Under
Setting One, where the external controls do not present any outcome
heterogeneity, all integrative estimators demonstrate improved Root-MSE
relative to the benchmark $\widehat{\theta}_{\tau}^{\text{aipw}}$.
Our proposal $\widehat{\theta}_{\tau}^{\text{adapt}}$ is less efficient
compared to other integrative estimators as it induces extra variability
due to bias detection for selective borrowing. However, $\widehat{\theta}_{\tau}^{\text{acw}}$,
$\widehat{\theta}^{\text{psrwe}}$ and $\widehat{\theta}_{\tau}^{\text{TransCox}}$
can be substantially biased under Settings Two and Three, where the
unmeasured confounder or time inconcurrency are present. In particular, our data-adaptive integrative estimator can detect the incompatibility
of external controls and selectively borrows the comparable subset,
resulting in a similar level of biases, but improved standard errors compared to the benchmark $\widehat{\theta}_{\tau}^{\text{aipw}}$. The results under Settings Four and Five are presented in Appendix \ref{sec:additional}.

\begin{figure}[!htbp]
\centering
\includegraphics[width=.6\linewidth]{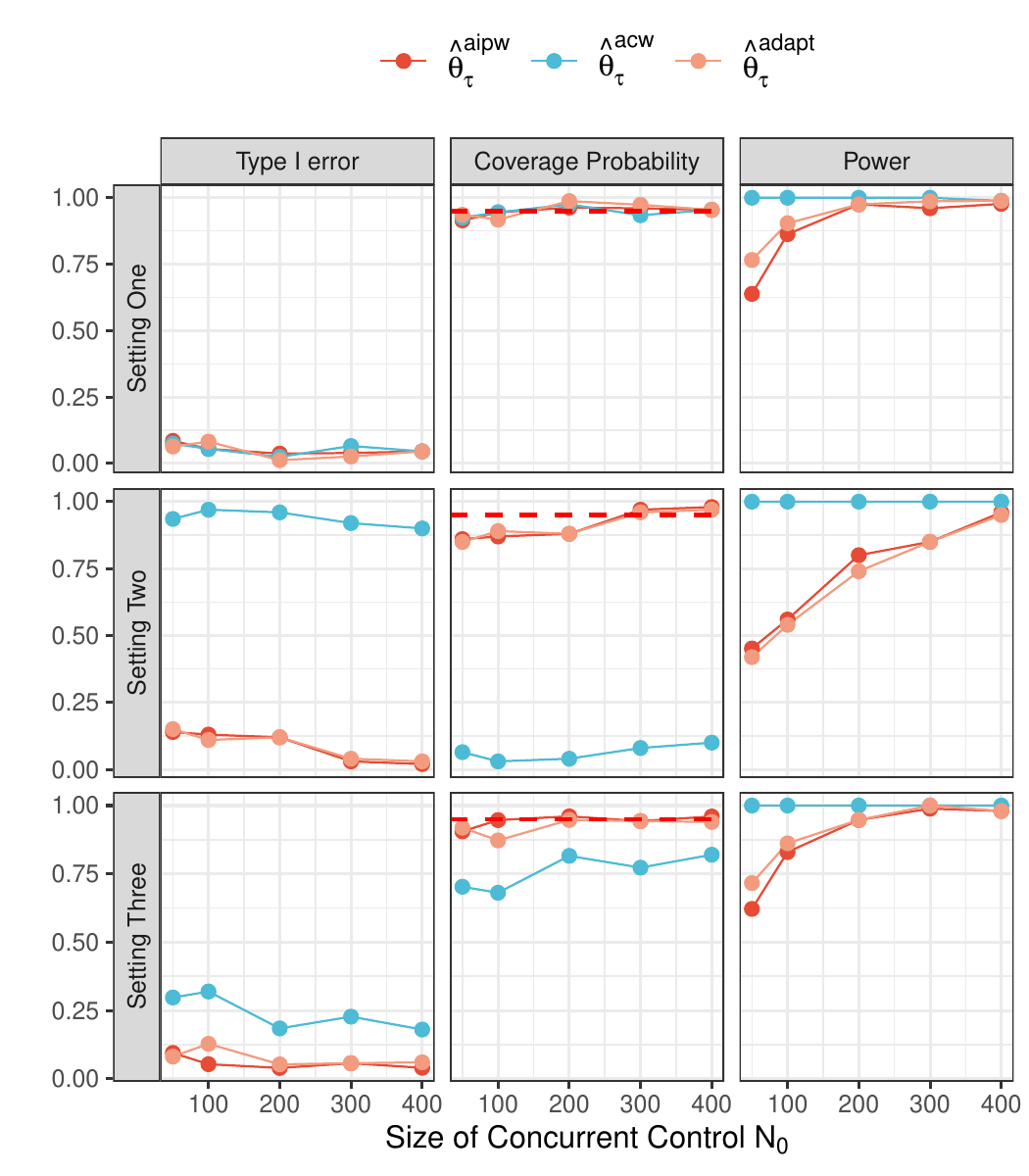}
\caption{\label{fig:sim-inference} Asymptotic results for RMST over $500$ Monte
Carlo experiments under Settings 1) selection bias only; 2) unmeasured confounding; 3) lack of concurrency.}
\end{figure}

To evaluate the asymptotic properties of our proposed estimators, Figure \ref{fig:sim-inference} presents the type-I error, the coverage probability,
and the power for detecting $\theta_{\tau}>-0.3$ for the estimators
$\widehat{\theta}_{\tau}^{\text{aipw}}$, $\widehat{\theta}_{\tau}^{\text{acw}}$,
and $\widehat{\theta}_{\tau}^{\text{adapt}}$. The bootstrap-based variance estimation with bootstrap size $50$ is used to construct the $95\%$ Wald confidence intervals for evaluation. Under Setting One,
$\widehat{\theta}_{\tau}^{\text{acw}}$ successfully controls the
type-I errors, maintain the nominal coverage rates and achieve the highest powers for detecting treatment effects over a varying size
of concurrent controls. However, it has inflated type-I errors and deteriorated coverage probabilities under Setting Two and Three when the outcome drift is present. In contrast, our proposed borrowing estimator effectively controls the type-I error, maintains
the satisfactory coverage rates, and achieves improved or comparable
power levels compared to the benchmark $\widehat{\theta}_{\tau}^{\text{aipw}}$,
irrespective of the presence of the outcome drift.

\section{Real-data Application}\label{sec:application}

This section presents an application of the proposed selective borrowing methodology to evaluate the effectiveness of Galcanezumab (120mg) versus placebo in patients with episodic migraine. The primary trial study is EVOLVE-1, a phase 3 double-blinded trial for patients with episodic migraines that randomized patients 1:1:2 to receive monthly Galcanezumab 120mg, 240mg, or placebo for up to 6 months \citep{stauffer2018evaluation}. In addition, placebo subjects from the REGAIN study were used as external controls to augment the control arm from the EVOLVE-1 study. The REGAIN study is a phase-3 double blinded trial for patients with chronic migraine headaches that randomized patients 1:1:2 to receive monthly galcanezulab 120mg, 240mg, or placebo for up to 3 months, with a subsequent 9-month open label extension follow-up period \citep{detke2018galcanezumab}.

The primary objective is to assess whether galcanezumab 120mg is superior to placebo in helping patients with episodic migraine achieve a meaningful improvement in migraine headache days (MHD), defined as a 50\% reduction in mean MHD per month from baseline. To estimate the difference in time to first meaningful MHD reduction up to 6 months post-baseline, the galcanezumab 120mg and placebo arms from EVOLVE-1 are augmented with the placebo arm from the REGAIN study. The treatment effect $\theta_{\tau}$ is defined as the RMST difference of the time to first occurrence of 50\% MHD reduction up to time $\tau = 6$ months.

\begin{figure}[!htbp]
\centering
\includegraphics[width=\linewidth]{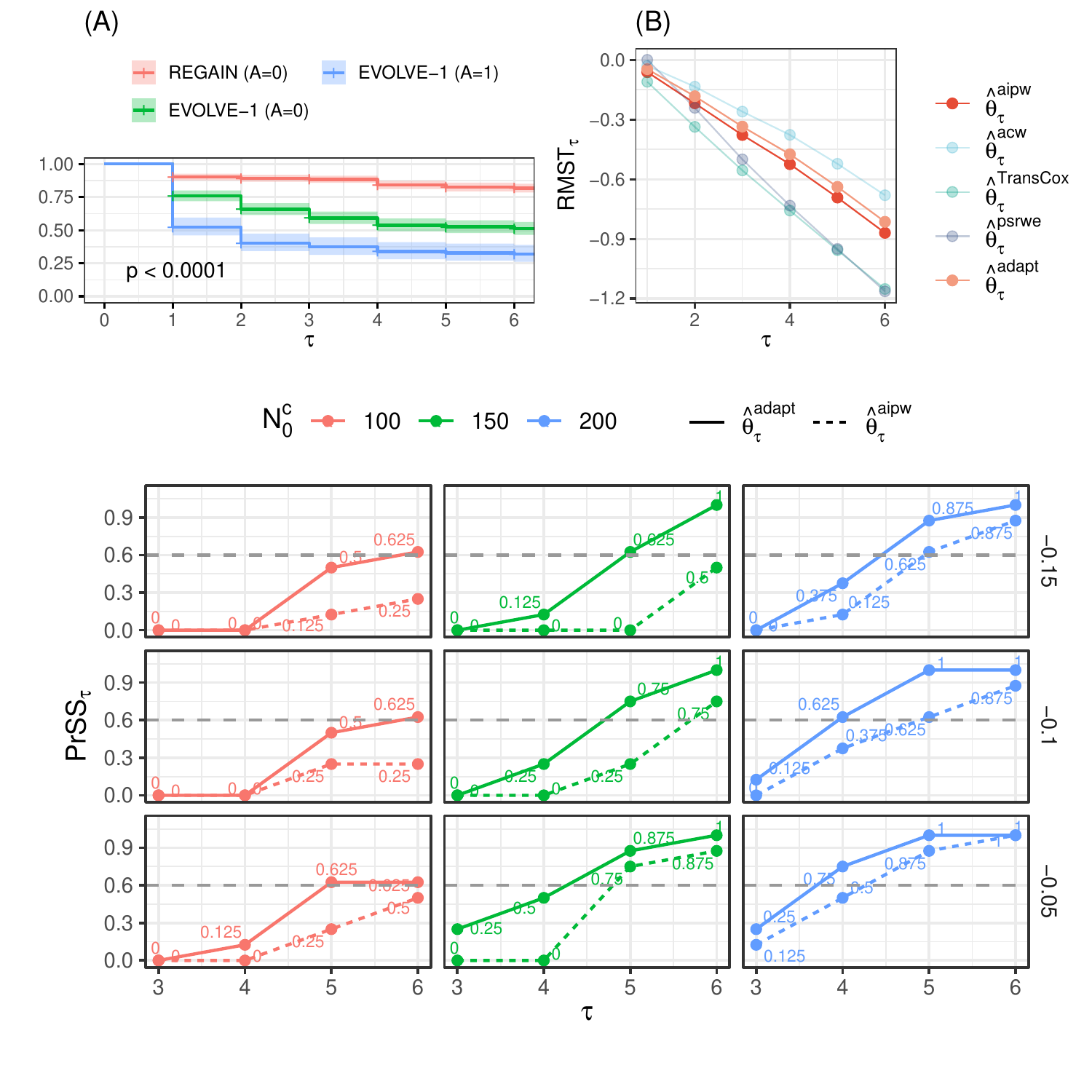}\caption{\label{fig:real-full} (A) Kaplan-Meier survival curves for the EVOLVE-1 study and the placebo group of the REGAIN study; (B) The estimated treatment effect in RMST for the EVOLVE-1 study; (C) Probability of study
success of detecting $\theta_{\tau}<-0.05, -0.1, -0.15$ using $\widehat{\theta}_{\tau}^{\text{aipw}}$
and $\widehat{\theta}_{\tau}^{\text{adapt}}$ over a range of restricted
times $\tau$ under varying sizes of sub-samples from the placebo group of the REGAIN study.}
\end{figure}

Figure \ref{fig:real-full}(A) presents the unconditional Kaplan-Meier
curves for the placebo group of these two studies. The log-rank test
indicates that the time to meaningful MHD improvement is different between the placebo arms of the EVOLVE-1 study and the REGAIN study, suggesting the need to account for outcome drift. Figure \ref{fig:real-full}(B) provides the estimated
treatment effect of Galcanezumab (120mg) in terms of RMST as a function
of the restricted month $\tau$. All estimators exhibit a trend of
decreasing RMST and show significant improvements in mitigating migraines
severity, implying a shortened response time for at least a $50\%$
reduction in MHD after being treated with Galcanezumab (120mg). Our
proposed estimator $\widehat{\theta}_{\tau}^{\text{adapt}}$yields
a RMST that is closer to the trial-only estimator $\widehat{\theta}_{\tau}^{\text{aipw}}$,
highlighting its capability to control the bias arising from external
heterogeneity. 

Next, we benchmark the performances of $\widehat{\theta}_{\tau}^{\text{adapt}}$
with the trial-only estimator $\widehat{\theta}_{\tau}^{\text{aipw}}$
to emphasize the efficiency gain of our selective borrowing framework.
To accomplish this, we retain the treatment group of the primary EVOLVE-1 study with a size of $N_{1}=212$, and create $50$ sub-samples by
randomly selecting $N_{0}^{c}$ patients from its placebo group with
$N_{0}^{c}=100,150,200$. The placebo group of the REGAIN study is then augmented to each selected sub-samples. Figure \ref{fig:real-full}(C)
presents the empirical probability of study success (PrSS) over a
range of restricted times $\tau$ under different sizes of concurrent
controls. The empirical PrSS is computed by the proportion of successfully
detecting $\theta_{\tau}<-0.05, -0.1, -0.15$ over the repeated sub-samples. When
combined with the placebo group of the REGAIN study, $\widehat{\theta}_{\tau}^{\text{adapt}}$
enhances the time-to-event analyses and yields a higher PrSS compared
to $\widehat{\theta}_{\tau}^{\text{aipw}}$across all the sizes of
sampled concurrent controls. 

For example, suppose that we aim to reach PrSS $\geq 0.6$ of detecting $\theta_{\tau}<-0.1$ at month $\tau=6$, $\widehat{\theta}_{\tau}^{\text{adapt}}$ only need to recruit 100 patients for the placebo group (solid red line at  $\tau = 6$ of the middle panel in Figure \ref{fig:real-full}(C)), however, the trial-only estimator $\widehat{\theta}_{\tau}^{\text{aipw}}$ needs at least 150 patients for the placebo group (dash green line at $\tau = 6$ of the middle panel in Figure \ref{fig:real-full}(C)). Therefore, our approach could attain similar levels of PrSS with fewer patients and a shortened patient enrollment
period by appropriately leveraging the external controls, which could eventually accelerate the drug development for rare diseases where
the event rates are typically low and imbalanced trials are often considered.

\section{Discussion}\label{sec:Discussion}

In this paper, we introduce a doubly protected borrowing framework that utilizes external controls to enhance treatment estimation for survival outcomes. Unlike most existing approaches, our method is built on semi-parametric efficient estimation coupled with the DR-Learner to detect outcome drift for selective borrowing. This approach effectively incorporates external controls without introducing biases into the integrative treatment evaluation. Moreover, the proposed approach offers a new perspective on the integrative analysis for survival outcomes with a proper method for inference, which could be a valuable contribution to many survival analyses in the machine learning community, such as those involving customer churn \citep{lariviere2004investigating,gao2024causal}, and multi-source domain adaption \citep{mansour2008domain, shaker2023multi}.

Our simulation studies reveal several challenges in controlling type I errors in the presence of unmeasured confounding for small samples, a problem also noted in other methods such as Bayesian dynamic borrowing \citep{dejardin2018use,kopp2020power}. Future work will focus on enhancing type I error control using alternative strategies, such as exact inference. In addition, the current selection criterion focuses on detecting outcome drift based on the differences in RMSTs. However, this may be less efficient when some parts of the conditional survival function are invariant across studies, as our proposal might exclude these valuable external controls that are partially comparable to the trial data. Future research should explore approaches that select comparable information rather than entire comparable subjects. One promising direction involves jointly estimating the average treatment effect and bias functions, as suggested in recent studies \citep{yang2020improved, wu2023transfer}.

\section*{Software and Data}
Our R codes with illustrative examples are available at \href{https://github.com/Gaochenyin/SelectiveIntegrative}{here}.

\section*{Acknowledgements}
We would like to thank the anonymous (meta-)reviewers of ICML 2025 for their helpful comments. This project is supported by the Food and Drug Administration (FDA) of the U.S. Department of Health and Human Services (HHS) as part of a financial assistance award U01FD007934 totaling $1,674,013$ over two years funded by FDA/HHS. It is also supported by the National Science Foundation under Award Number SES 2242776, totaling $225,000$ over three years. The contents are those of the authors and do not necessarily represent the official views of, nor an endorsement by, FDA/HHS, the National Science Foundation, or the U.S. Government.

\section*{Impact Statement}
Our proposed method for analyzing censored survival data enhances the evaluation of treatment efficacy in clinical trials with small control groups by incorporating external controls while addressing data heterogeneity to reduce bias. This approach  promotes ethical research practices by improving statistical efficiency and minimizing patient exposure to ineffective treatments.

\bibliographystyle{apalike}
\bibliography{main}

\newpage
\appendix
\section*{Appendix}

\renewcommand\thefigure{\thesection.\arabic{figure}}
\renewcommand\thetable{\thesection.\arabic{table}}
\section{Proofs}\label{sec:proofs}
\subsection{Proof of Theorem \ref{thm:identification0}}\label{subsec:proof_identification0}

We only prove the identification formulas for $S_{0}(t\mid R=1)$,
and similar proofs follow for $S_{1}(t\mid R=1)$. Based on the trial
data, we have 
\begin{align}
S_{0}(t\mid R=1) & =\E\{S_{0}(t\mid X,R=1)\mid R=1\}\nonumber \\
 & =\frac{1}{\pr(R=1)}\E\{RS_{0}(t\mid X,R=1)\}\label{eq:identi}\\
 & =\frac{1}{\pr(R=1)}\E\{RP(T^{(0)}>t\mid X,R=1)\}\nonumber \\
 & =\frac{1}{\pr(R=1)}\E\{RP(T^{(0)}>t\mid X,R=1,A=0,C>T)\}\nonumber \\
 & =\frac{1}{\pr(R=1)}\E\left\{ \frac{R(1-A)\Delta\mathbf{1}(Y>t)}{\{1-\pi_{A}(X)\}S^{C}(Y\mid X,A=0,R=1)}\right\} ,\nonumber 
\end{align}
where the fourth equality holds under Assumptions \ref{assum:trial}
and \ref{assum:censoring_noninformative}. Based on the external data,
we have 
\begin{align*}
S_{0}(t\mid R=1) & =\E\{S_{0}(t\mid X,R=1)\mid R=1\}\\
 & =\frac{1}{\pr(R=1)}\E\{RS_{0}(t\mid X,R=1)\}\\
 & =\frac{1}{\pr(R=1)}\E\{\pi_{R}(X)S_{0}(t\mid X,R=0)\}\\
 & =\frac{1}{\pr(R=1)}\E\{\pi_{R}(X)P(T^{(0)}>t\mid X,R=0,A=0,C>T)\}\\
 & =\frac{1}{\pr(R=1)}\E\left\{ \frac{(1-R)q_{R}(X)\Delta\mathbf{1}(Y>t)}{S^{C}(Y\mid X,A,R=0)}\right\} ,
\end{align*}
where the third equality holds under Assumptions \ref{assum:exchangeability_mean}.

\subsection{Proof of Theorem \ref{thm:EIF0}}\label{sec:proof_EIF0}

\subsubsection{Preliminaries}

We first derive the full-data efficient influence function (EIF) for
the treatment-specific survival curves $S_{a}(t\mid R=1)$ without
censoring, i.e., the full data $W_{i}=(T_{i},A_{i},X_{i},R_{i}=r)$.
We next employ the semi-parametric theory in \citet{bickel1993efficient}
to derive the EIFs. In particular, we consider a one-dimensional parametric
submodel $f_{\theta}(W)$, which contains the true model $f(W)$ at
$\theta=0$, i.e., $f_{\theta}(W)\mid_{\theta=0}=f(W)$. We use dot
to denote the partial derivative with respect to $\theta$, and $s_{\theta}(\cdot)$
to denote the score function of the submodel. For example, we have
\begin{align*}
\dot{\mu}_{0} & =\frac{\partial}{\partial\theta}\E_{\theta}\{\mu_{0}(X)\}=\int_{\mathcal{X}}\mu_{0}(X)\frac{\partial f_{\theta}(X)}{\partial X}dX\\
 & =\int_{\mathcal{X}}\mu_{0}(X)\frac{\dot{f}_{\theta}(X)}{f_{\theta}(X)}f_{\theta}(X)dX=\E\{\mu_{0}(X)s_{\theta}(X)\},
\end{align*}
where $s_{\theta}(X)=\partial\log f_{\theta}(X)/\partial\theta$.
Further, we can factorize the full-data likelihood function as:
\begin{align*}
f(W) & =f(X)\pr(R=1\mid X)^{R}\pr(R=0\mid X)^{1-R}\\
 & \times\pr(A=1\mid X,R=1)^{RA}\pr(A=0\mid X,R=1)^{R(1-A)}\\
 & \times f(T\mid X,R=1,A=1)^{RA}f(T\mid X,R=1,A=0)^{R(1-A)}\\
 & \times f(T\mid X,R=0)^{1-R},
\end{align*}
and the associated score function under the submodel can be decomposed
as
\begin{align*}
s_{\theta}(W) & =s_{\theta}(X)+\frac{R-\pr(R=1\mid X)}{\pr(R=1\mid X)\{1-\pr(R=1\mid X)\}}\dot{\pr}_{\theta}(R=1\mid X)\\
 & +\frac{R\{A-\pi_{A}(X)\}}{\pi_{A}(X)\{1-\pi_{A}(X)\}}\dot{\pr}_{\theta}(A=1\mid X,R=1)\\
 & +RAs_{\theta}(T\mid X,A=1,R=1)+R(1-A)s_{\theta}(T\mid X,R=1,A=0)\\
 & +(1-R)s_{\theta}(T\mid X,R=0),
\end{align*}
where $s_{\theta}(X)=\partial\log f_{\theta}(X)/\partial\theta$,
$s_{\theta}(T\mid X,R=1,A=a)=\partial\log f_{\theta}(T\mid X,R=1,A=a)/\partial\theta$
for $a=0,1$, and $s_{\theta}(T\mid X,R=0)=\partial\log f_{\theta}(T\mid X,R=0)/\partial\theta$.
Analogous to our definition $f_{\theta}(W)\mid_{\theta=0}=f(W)$,
we have $s_{\theta}(\cdot)\mid_{\theta=0}=s(\cdot)$, which is the
true score function evaluated at the true parameter under the one-dimensional
submodel.

\subsubsection{Full-data Efficient Influence Function}

From the semiparametric theory, the orthogonal complement of the full-data
nuisance tangent space $\Lambda_{F}^{\perp}$ equals to
\begin{equation}
\Lambda_{F}^{\perp}=H_{1}\oplus H_{2}\oplus H_{3}\oplus H_{4},\label{eq:tangent_space}
\end{equation}
where 
\begin{align*}
H_{1} & =\{\Gamma(X):\E\{\Gamma(X)\}=0\},\\
H_{2} & =\left\{ \{R-\pr(R=1\mid X)\}a(X)\right\} ,\quad H_{3}=\left\{ R\{A-\pi_{A}(X)\}b(X)\right\} ,\\
H_{4} & =H_{41}\cap H_{42}=\left\{ \Gamma(T,X,R,A):\E\{\Gamma(T,X,R,A)\mid X,R,A\}=0\right\} \\
 & \cap\left\{ \Gamma(T,X,R,A):\E\left[\left\{ \frac{(1-R)\mathbf{1}(T>t)}{P(R=0\mid X)}-\frac{R(1-A)\mathbf{1}(T>t)}{P(R=1,A=0\mid X)}\right\} \Gamma(T,X,R,A)\mid X\right]=0,t<\tau\right\} ,
\end{align*}
for any two arbitrary square-integrable measurable functions $a(X)$
and $b(X)$. The tangent space $H_{42}$ is induced by the conditional
mean exchangeability in Assumption \ref{assum:exchangeability_mean},
where $S_{0}(t\mid X,R=1)=S_{0}(t\mid X,R=0)$ for any $t<\tau$.
The EIF for $S_{0}(t\mid R=1)$, denoted by $\psi_{S_{0},\eff}^{F}(t,W)\in\Lambda_{F}^{\perp}$
should satisfy 
\[
\partial S_{0}(t\mid R=1)/\partial\theta\mid_{\theta=0}=\E\{\psi_{S_{0},\eff}^{F}(t,W)s(W)\}.
\]
Based on our identification formula (\ref{eq:identi}), $S_{0}(t\mid R=1)=\E\{\pi_{R}(X)S_{0}(t\mid X,R=1)\}/\pr(R=1)$,
which is a ratio for with numerator $N=\E\{\pi_{R}(X)S_{0}(t\mid X,R=1)\}$
and denominator $D=\pr(R=1)$. Therefore, our strategy is to first
derive the EIF for the numerator and denominator, and then combine
them to have the final full-data EIF $\psi_{S_{0},\eff}^{F}(t,W)$.

Let $N_{\theta}$ and $D_{\theta}$ denote $N$ and $D$ being evaluated
at the submodel $f_{\theta}(W)$. For the numerator, pathwise derivative
is evaluated by the chain rule:
\begin{align}
\frac{\partial N_{\theta}}{\partial\theta}\mid_{\theta=0} & =\E\{\pr(R=1\mid X)\pr(T>t\mid R=1,A=0,X)s(X)\}\nonumber \\
 & +\E\left\{ \frac{\partial\pr_{\theta}(R=1\mid X)}{\partial\theta}\pr(T>t\mid R=1,A=0,X)\right\} \mid_{\theta=0}\label{eq:EIF_N_second}\\
 & +\E\left\{ \pr(R=1\mid X)\frac{\partial\pr_{\theta}(T>t\mid R=1,A=0,X)}{\partial\theta}\right\} \mid_{\theta=0}.\label{eq:EIF_N_third}
\end{align}
Next, we show the second part of the pathwise derivative (\ref{eq:EIF_N_second})
is 
\[
\frac{\partial\pr_{\theta}(R=1\mid X)}{\partial\theta}=\E[\{R-\pr(R=1\mid X)\}s(A,R\mid X)].
\]
However, the pathwise derivative $\partial\pr_{\theta}(T>t\mid X,R=1)/\partial\theta$ in the third part can be derived in different ways under Assumption
\ref{assum:exchangeability_mean} with the trial data by (\ref{eq:EIF-trial})
and the external controls by (\ref{eq:EIF-control}): 
\begin{align}
 & \frac{\partial}{\partial\theta}\pr_{\theta}(T>t\mid R=1,A=0,X)\nonumber \\
 & =\frac{\partial}{\partial\theta}\E_{\theta}\left\{ \mathbf{1}(T>t)\mid R=1,A=0,X\right\} \nonumber \\
 & =\int_{t}^{\infty}\mathbf{1}(T>t)\frac{\partial}{\partial\theta}f_{\theta}(T\mid R=1,A=0,X)dT\nonumber \\
 & =\int_{t}^{\infty}\mathbf{1}(T>t)s_{\theta}(T\mid R=1,A=0,X)f_{\theta}(T\mid R=1,A=0,X)dT\nonumber \\
 & =\int_{t}^{\infty}\frac{R(1-A)f_{\theta}(T\mid X)\left\{ \mathbf{1}(T>t)-S_{0}(t\mid X,R=1)\right\} }{\pr(R=1,A=0\mid X)}s_{\theta}(T\mid R,A,X)dT\nonumber \\
 & =\E\left[\frac{R(1-A)\left\{ \mathbf{1}(T>t)-S_{0}(t\mid X,R=1)\right\} s_{\theta}(T\mid R,A,X)}{\pr(R=1,A=0\mid X)}\mid X\right],\label{eq:EIF-trial}
\end{align}
and 
\begin{align}
 & \frac{\partial}{\partial\theta}\pr_{\theta}(T>t\mid R=1,A=0,X)\nonumber \\
 & =\frac{\partial}{\partial\theta}\pr_{\theta}(T>t\mid R=0,A=0,X)\nonumber \\
 & =\frac{\partial}{\partial\theta}\E_{\theta}\left\{ \mathbf{1}(T>t)\mid R=0,A=0,X\right\} \nonumber \\
 & =\int_{t}^{\infty}\mathbf{1}(T>t)\frac{\partial}{\partial\theta}f_{\theta}(T\mid R=0,A=0,X)dT^{(0)}\nonumber \\
 & =\int_{t}^{\infty}\mathbf{1}(T>t)s_{\theta}(T\mid R=0,A=0,X)f_{\theta}(T\mid R=0,A=0,X)dT\nonumber \\
 & =\int_{t}^{\infty}\left\{ \mathbf{1}(T>t)-S_{0}(t\mid X,R=0)\right\} s_{\theta}(T\mid X,R)\frac{(1-R)f_{\theta}(T\mid X)}{\pr(R=0\mid X)}dT\nonumber \\
 & =\E\left[\frac{(1-R)\left\{ \mathbf{1}(T>t)-S_{0}(t\mid X,R=0)\right\} s_{\theta}(T\mid X,R)}{\pr(R=0\mid X)}\mid X\right].\label{eq:EIF-control}
\end{align}
To obtain the efficient influence function of $N$, we need to find
the proper functions $C_{1}$ and $C_{2}$ of $(X,R,A)$ such that
the third part (\ref{eq:EIF_N_third}) belongs to the tangent space
$H_{4}$, which satisfies: 
\begin{align}
\E & \left(\left[C_{1}(1-R)q(X)\left\{ \mathbf{1}(T>t)-S_{0}(t\mid X,R=1)\right\} +C_{2}R(1-A)\frac{\mathbf{1}(T>t)-S_{0}(t\mid X,R=1)}{1-\pi_{A}(X)}\right]\right.\label{eq:C1_C2}\\
 & \left.\times\left\{ \frac{(1-R)\mathbf{1}(T>t)}{\pr(R=0\mid X)}-\frac{R(1-A)\mathbf{1}(T>t)}{\pr(R=1,A=0\mid X)}\right\} \mid X\right)=0.\nonumber 
\end{align}
By simple algebra, we can show that 
\begin{align*}
\frac{C_{1}}{C_{2}} & =\frac{\E\left[R^{2}(1-A)^{2}\{\mathbf{1}(T>t)-S_{0}(t\mid X,R=1)\}\mathbf{1}(T>t)\mid X\right]}{\{1-\pi_{A}(X)\}\pr(R=1,A=0\mid X)}\\
 & \left(\frac{\E\left[(1-R)^{2}q_{R}(X)\{\mathbf{1}(T>t)-S_{0}(t\mid X,R=1)\}\mathbf{1}(T>t)\mid X\right]}{P(R=0\mid X)}\right)^{-1}\\
 & =\frac{r(t,X)}{\{1-\pi_{A}(X)\}q_{R}(X)},
\end{align*}
where $r(t,X)=\text{var}\left\{ \mathbf{1}(T>t)\mid R=1,A=0\right\} /\text{var}\left\{ \mathbf{1}(T>t)\mid R=0\right\} $.
Plugging the above formulas, we obtain the EIF for the numerator $N$
as: 
\begin{align*}
\psi_{N,\eff}^{F}(t,W) & =RS_{0}(t\mid X,R=1)\\
 & +\frac{(1-R)r(t,X)q_{R}(X)\left\{ \mathbf{1}(T>t)-S_{0}(t\mid X,R=1)\right\} }{D(t,X)}\\
 & +\frac{R(1-A)q_{R}(X)\left\{ \mathbf{1}(T>t)-S_{0}(t\mid X,R=1)\right\} }{D(t,X)}.
\end{align*}
where $D(t,X)=r(t,X)+\{1-\pi_{A}(X)\}q_{R}(X)$. For the denominator
$D_{\theta}$, we have $\partial D_{\theta}/\partial\theta\mid_{\theta=0}=\E\{R\partial\pr_{\theta}(R=1)/\partial\theta\}\mid_{\theta=0}$.
By Lemma S1 in \citet{jiang2022multiply}, the full-data EIF of $N/D$
is obtained by:
\begin{align*}
\psi_{S_{0},\eff}^{F}(t,W) & =\frac{\psi_{N,\eff}^{F}(t,W)-RS_{0}(t\mid R=1)}{\pr(R=1)}\\
 & =\frac{R(1-A)}{\pr(R=1)}\frac{q_{R}(X)\left\{ \mathbf{1}(T>t)-S_{0}(t\mid X,R=1)\right\} }{D(t,X)}\\
 & +\frac{1-R}{\pr(R=1)}\frac{r(t,X)q_{R}(X)\left\{ \mathbf{1}(T>t)-S_{0}(t\mid X,R=1)\right\} }{D(t,X)}\\
 & +\frac{R}{\pr(R=1)}\{S_{0}(t\mid X,R=1)-S_{0}(t\mid R=1)\},\\
 & =\frac{R(1-A)}{\pr(R=1)}\frac{q_{R}(X)\left\{ \mathbf{1}(T>t)-S_{0}(t\mid R=1)\right\} }{D(t,X)}\\
 & +\frac{(1-R)q_{R}(X)}{\pr(R=1)}\frac{r(t,X)\left\{ \mathbf{1}(T>t)-S_{0}(t\mid R=1)\right\} }{D(t,X)}\\
 & +\frac{S_{0}(t\mid X,R=1)-S_{0}(t\mid R=1)}{\pr(R=1)}\left\{ \frac{R\{A-\pi_{A}(X)\}q_{R}(X)}{D(t,X)}+\frac{r(t,X)\{R-(1-R)q_{R}(X)\}}{D(t,X)}\right\} ,
\end{align*}
which belongs to the tangent space $\Lambda_{F}^{\perp}$.

\subsubsection{Observed-data Efficient Influence Function \label{subsec:observed-EIF}}

In the presence of censoring, i.e., a special form of monotone coarsening,
we observe the data set $V_{i}=(Y_{i},\Delta_{i},A_{i},X_{i},R_{i}=r)$
instead of $W_{i}$. Define the many-to-one linear mapping $\mathcal{K}:\Lambda_{\eta}^{\perp}\rightarrow\Lambda_{F}^{\perp}$
to be $\mathcal{K}(h)=\E\{h(V)\mid W\}$ for any $h\in\Lambda_{\eta}^{\perp}$,
where $\Lambda_{\eta}^{\perp}$ is the orthogonal complement of the
observed-data nuisance tangent space by Lemma 7.3, \citet{tsiatis2006semiparametric}.
Let $\psi^{F}(t,W)$ be a typical element of $\Lambda_{F}^{\perp}$,
by Theorem 7.2 from \citet{tsiatis2006semiparametric}, the space
$\Lambda_{\eta}^{\perp}$ consists of all elements that can be written
as
\begin{align}
\Lambda_{\eta}^{\perp}=\mathcal{K}^{-1}(\Lambda_{F}^{\perp}) & =\frac{\Delta\psi^{F}(t,W)}{\pr(\Delta=1\mid W)}+\mathcal{K}^{-1}(0)\label{eq:observed-Lambda}
\end{align}
where $\mathcal{K}^{-1}$ as the inverse operator, and $\mathcal{K}^{-1}(0)$
consists any arbitrary functions $L(t,V)$ such that $\E\{L(t,V)\mid W\}=0$.
The first part of (\ref{eq:observed-Lambda}) is motivated by the
inverse censoring weighting of the complete-case estimator where the
event time is observed, indicated by $\Delta=1$. The second part
of (\ref{eq:observed-Lambda}) is referred to as the augmentation
space due to censoring. The optimal element (\ref{eq:observed-Lambda})
with the greatest efficiency improvement is obtained by projecting
$\Delta\psi^{F}(t,W)/\pr(\Delta=1\mid W)$ to the tangent space $\mathcal{K}^{-1}(0)$.
By Theorems 9.2 and 10.4, we derive the observed-data EIF for $S_{0}(t\mid R=1)$
under our monotone coarsened data: 
\begin{align}
 & \psi_{S_{0},\eff}(t,V)=\frac{R(1-A)\Delta}{\pr(R=1)S^{C}(Y\mid X,R=1)}\frac{q_{R}(X)\left\{ \mathbf{1}(Y>t)-S_{0}(t\mid R=1)\right\} }{D(t,X)}\label{eq:IPW1}\\
 & +\frac{(1-R)\Delta q_{R}(X)}{\pr(R=1)S^{C}(Y\mid X,R=0)}\frac{r(t,X)\left\{ \mathbf{1}(Y>t)-S_{0}(t\mid R=1)\right\} }{D(t,X)}\label{eq:IPW2}\\
 & +\frac{S_{0}(t\mid X,R=1)-S_{0}(t\mid R=1)}{\pr(R=1)}\left\{ \frac{R\{A-\pi_{A}(X)\}q_{R}(X)}{D(t,X)}+\frac{r(t,X)\{R-(1-R)q_{R}(X)\}}{D(t,X)}\right\} \label{eq:aug_part_RA}\\
 & +\int_{0}^{\infty}\frac{R(1-A)}{\pr(R=1)}\frac{dM_{0}^{C}(r\mid X)}{S^{C}(r\mid X,R=1)}\E\left[\frac{q_{R}(X)\left\{ \mathbf{1}(T>t)-S_{0}(t\mid R=1)\right\} }{D(t,X)}\mid T>r,X\right]\label{eq:aug_part1}\\
 & +\int_{0}^{\infty}\frac{(1-R)q_{R}(X)}{\pr(R=1)}\frac{dM_{0}^{C}(r\mid X)}{S^{C}(r\mid X,R=0)}\E\left[\frac{r(t,X)\left\{ \mathbf{1}(T>t)-S_{0}(t\mid R=1)\right\} }{D(t,X)}\mid T>r,X\right],\label{eq:aug_part2}
\end{align}
where $dM_{0}^{C}(r\mid X)=dN_{0}^{C}(r)-\mathbf{1}(Y\geq r)\lambda_{0}^{C}(r\mid X)$.
The last two terms (\ref{eq:aug_part1}) and (\ref{eq:aug_part2})
belong to the augmentation space $\mathcal{K}^{-1}(0)$. Next, we
can simplify it to be the EIF in Theorem \ref{thm:EIF0}. Note that
(\ref{eq:aug_part1}) can be expressed by
\begin{align}
 & \int_{0}^{\infty}\frac{R(1-A)}{\pr(R=1)}\frac{dM_{0}^{C}(r\mid X)}{S^{C}(r\mid X,R=1)}\E\left[\frac{q_{R}(X)\left\{ \mathbf{1}(T>t)-S_{0}(t\mid R=1)\right\} }{D(t,X)}\mid T>r,X\right]\nonumber \\
= & \int_{0}^{\infty}\frac{R(1-A)}{\pr(R=1)}\frac{dM_{0}^{C}(r\mid X)q_{R}(X)}{S^{C}(r\mid X,R=1)D(t,X)}\mathbf{1}(r<t)\left\{ \frac{S_{0}(t\mid X,R=1)}{S_{0}(r\mid X,R=1)}-S_{0}(t\mid R=1)\right\} \nonumber \\
 & +\int_{0}^{\infty}\frac{R(1-A)}{\pr(R=1)}\frac{dM_{0}^{C}(r\mid X)q_{R}(X)}{S^{C}(r\mid X,R=1)D(t,X)}\mathbf{1}(r\geq t)\left\{ 1-S_{0}(t\mid R=1)\right\} \nonumber \\
= & \int_{0}^{t}\frac{R(1-A)}{\pr(R=1)}\frac{dM_{0}^{C}(r\mid X)q_{R}(X)}{S^{C}(r\mid X,R=1)D(t,X)}\frac{S_{0}(t\mid X,R=1)}{S_{0}(r\mid X,R=1)}\label{eq:term1}\\
 & +\int_{t}^{\infty}\frac{R(1-A)}{\pr(R=1)}\frac{dM_{0}^{C}(r\mid X)q_{R}(X)}{S^{C}(r\mid X,R=1)D(t,X)}\label{eq:term2}\\
 & -S_{0}(t\mid R=1)\int_{0}^{\infty}\frac{R(1-A)}{\pr(R=1)}\frac{dM_{0}^{C}(r\mid X)q_{R}(X)}{S^{C}(r\mid X,R=1)D(t,X)}.\label{eq:term3}
\end{align}
The second term (\ref{eq:term2}) equals to 
\begin{align*}
 & \int_{t}^{\infty}\frac{R(1-A)}{\pr(R=1)}\frac{dM_{0}^{C}(r\mid X)q_{R}(X)}{S^{C}(r\mid X,R=1)D(t,X)}\\
 & =\frac{R(1-A)q_{R}(X)}{\pr(R=1)D(t,X)}\int_{t}^{\infty}\frac{dM_{0}^{C}(r\mid X)}{S^{C}(r\mid X,R=1)}\\
 & =\frac{R(1-A)q_{R}(X)}{\pr(R=1)D(t,X)}\int_{t}^{\infty}\frac{dN_{0}^{C}(r)-\mathbf{1}(Y\geq r)\lambda_{0}^{C}(r\mid X)}{S^{C}(r\mid X,R=1)}\\
 & =\frac{R(1-A)q_{R}(X)\mathbf{1}(Y\geq t)}{\pr(R=1)D(t,X)}\left\{ \frac{1-\Delta}{S^{C}(Y\mid X,R=1)}-\int_{t}^{Y}\frac{\lambda_{0}^{C}(r\mid X)dr}{S^{C}(r\mid X,R=1)}\right\} \\
 & =\frac{R(1-A)q_{R}(X)\mathbf{1}(Y\geq t)}{\pr(R=1)D(t,X)}\left\{ \frac{1}{S^{C}(t\mid X,R=1)}-\frac{\Delta}{S^{C}(Y\mid X,R=1)}\right\} ,
\end{align*}
where the last equality holds as we know that $\lambda_{0}^{C}(r\mid X)=-\partial\log S^{C}(r\mid X,R=1)/\partial r$,
and $\int\lambda_{0}^{C}(r\mid X)/S^{C}(r\mid X,R=1)=1/S^{C}(r\mid X,R=1)$.
For the third term (\ref{eq:term3}), we have 
\begin{align*}
 & S_{0}(t\mid R=1)\int_{0}^{\infty}\frac{R(1-A)}{\pr(R=1)}\frac{dM_{0}^{C}(r\mid X)q_{R}(X)}{S^{C}(r\mid X,R=1)D(t,X)}\\
 & =S_{0}(t\mid R=1)\frac{R(1-A)q_{R}(X)}{\pr(R=1)D(t,X)}\int_{0}^{\infty}\frac{dN_{0}^{C}(r)-\mathbf{1}(Y\geq r)\lambda_{0}^{C}(r\mid X)}{S^{C}(r\mid X,R=1)}\\
 & =S_{0}(t\mid R=1)\frac{R(1-A)q_{R}(X)}{\pr(R=1)D(t,X)}\left\{ 1-\frac{\Delta}{S^{C}(Y\mid X,R=1)}\right\} .
\end{align*}
Plugging these formulas back with (\ref{eq:IPW1}) and (\ref{eq:aug_part1}),
we obtain
\begin{align*}
 & \frac{R(1-A)\Delta}{\pr(R=1)S^{C}(Y\mid X,R=1)}\frac{q_{R}(X)\left\{ \mathbf{1}(Y>t)-S_{0}(t\mid R=1)\right\} }{D(t,X)}\\
 & +\int_{0}^{t}\frac{R(1-A)}{\pr(R=1)}\frac{dM_{0}^{C}(r\mid X)q_{R}(X)}{S^{C}(r\mid X,R=1)D(t,X)}\frac{S_{0}(t\mid X,R=1)}{S_{0}(r\mid X,R=1)}\\
 & +\frac{R(1-A)q_{R}(X)\mathbf{1}(Y\geq t)}{\pr(R=1)D(t,X)}\left\{ \frac{1}{S^{C}(t\mid X,R=1)}-\frac{\Delta}{S^{C}(Y\mid X,R=1)}\right\} \\
 & -S_{0}(t\mid R=1)\frac{R(1-A)q_{R}(X)}{\pr(R=1)D(t,X)}\left\{ 1-\frac{\Delta}{S^{C}(Y\mid X,R=1)}\right\} \\
 & =\frac{R(1-A)q_{R}(X)\mathbf{1}(Y\geq t)}{\pr(R=1)D(t,X)S^{C}(t\mid X,R=1)}-\frac{R(1-A)q_{R}(X)S_{0}(t\mid R=1)}{\pr(R=1)D(t,X)}\\
 & +\int_{0}^{t}\frac{R(1-A)}{\pr(R=1)}\frac{dM_{0}^{C}(r\mid X)q_{R}(X)}{S^{C}(r\mid X,R=1)D(t,X)}\frac{S_{0}(t\mid X,R=1)}{S_{0}(r\mid X,R=1)}.
\end{align*}
Analogous simplification applies to the combination of (\ref{eq:IPW2})
and (\ref{eq:aug_part2}), which leads to 
\begin{align*}
 & \frac{(1-R)q_{R}(X)r(t,X)\mathbf{1}(Y>t)}{\pr(R=1)D(t,X)S^{C}(t\mid X,R=0)}-\frac{(1-R)q_{R}(X)r(t,X)S_{0}(t\mid R=1)}{\pr(R=1)D(t,X)}\\
 & +\int_{0}^{t}\frac{(1-R)}{\pr(R=1)}\frac{dM_{0}^{C}(r\mid X)q_{R}(X)r(t,X)}{S^{C}(r\mid X,R=1)D(t,X)}\frac{S_{0}(t\mid X,R=1)}{S_{0}(r\mid X,R=1)}.
\end{align*}
So the final observed-data EIF now becomes:
\begin{align*}
\psi_{S_{0},\eff}(t,V) & =\frac{R(1-A)q_{R}(X)\mathbf{1}(Y\geq t)}{\pr(R=1)D(t,X)S^{C}(t\mid X,R=1)}\\
 & +\int_{0}^{t}\frac{R(1-A)}{\pr(R=1)}\frac{dM_{0}^{C}(r\mid X)q_{R}(X)}{S^{C}(r\mid X,R=1)D(t,X)}\frac{S_{0}(t\mid X,R=1)}{S_{0}(r\mid X,R=1)}\\
 & +\frac{(1-R)q_{R}(X)r(t,X)\mathbf{1}(Y>t)}{\pr(R=1)D(t,X)S^{C}(t\mid X,R=0)}\\
 & +\int_{0}^{t}\frac{(1-R)}{\pr(R=1)}\frac{dM_{0}^{C}(r\mid X)q_{R}(X)r(t,X)}{S^{C}(r\mid X,R=1)D(t,X)}\frac{S_{0}(t\mid X,R=1)}{S_{0}(r\mid X,R=1)}\\
 & +\frac{S_{0}(t\mid X,R=1)}{\pr(R=1)}\left\{ \frac{R\{A-\pi_{A}(X)\}q_{R}(X)}{D(t,X)}+\frac{r(t,X)\{R-(1-R)q_{R}(X)\}}{D(t,X)}\right\} \\
 & -\frac{RS_{0}(t\mid R=1)}{\pr(R=1)},
\end{align*}
where 
\begin{align*}
\frac{R\{A-\pi_{A}(X)\}q_{R}(X)+R(1-A)q_{R}(X)}{\pr(R=1)D(t,X)} & =\frac{R\{1-\pi_{A}(X)\}q_{R}(X)}{\pr(R=1)D(t,X)},\\
\frac{r(t,X)\{R-(1-R)q_{R}(X)\}+(1-R)q_{R}(X)r(t,X)}{\pr(R=1)D(t,X)} & =\frac{Rr(t,X)}{\pr(R=1)D(t,X)}.
\end{align*}

\subsection{Proof of Theorem \ref{thm:theta_dr}}\label{sec:prrof_dr}

\begin{assumption}\label{assump:rate}
    Assume (a) $\|\widehat{\pi}_{R}(X)-\pi_{R}(X)\|_{L_{2}}=o_{\pr}(1)$, $\|\widehat{\pi}_{A}(X)-\pi_{A}(X)\|_{L_{2}}=o_{\pr}(1)$,
$\|\widehat{S}_{a}(t\mid X,R=1)-S_{a}(t\mid X,R=1)\|_{L_{2}}=o_{\pr}(1)$,
and $\|\widehat{S}^{C}(t\mid X,R)-S^{C}(t\mid X,R)\|_{L_{2}}=o_{\pr}(1)$; $0<c_{1}\leq\pi_{R}(X),\pi_{A}(X),S^{C}(t\mid X,R),S_{0}(t\mid X,R)\leq c_{2}<1$
and their estimated counterparts are bounded away from $0$ and $1$
for some constants $c_{1}$ and $c_{2}$; (c) $\|\widehat{S}_{a}(t\mid X,R=1)-S_{a}(t\mid X,R=1)\|_{L_{2}}\cdot\|\widehat{\pi}_{A}(X)-\pi_{A}(X)\|_{L_{2}}=o_{\pr}(N^{-1/2})$,
$\|\widehat{S}_{a}(t\mid X,R=1)-S_{a}(t\mid X,R=1)\|_{L_{2}}\cdot\|\widehat{\pi}_{R}(X)-\pi_{R}(X)\|_{L_{2}}=o_{\pr}(N^{-1/2})$,
and $\|\widehat{S}_{a}(t\mid X,R=1)-S_{a}(t\mid X,R=1)\|_{L_{2}}\cdot\max_{r<t}\|\widehat{\lambda}_{0}^{C}(r\mid X,R)-\lambda_{0}^{C}(r\mid X,R)\|_{L_{2}}=o_{\pr}(N^{-1/2})$ for
$a\in\{0,1\}$.
\end{assumption}
Assumption \ref{assump:rate} is analogous to those for double machine learning estimation for average treatment effects \citep{kennedy2016semiparametric}. To investigate the asymptotic properties of $\widehat{\theta}_{\tau}$,
we need to understand the components that constitute $\widehat{\theta}_{\tau}-\theta_{\tau}$.
Let $\mathbb{P}_{N}$ be the empirical measure, we have 
\begin{align}
\widehat{\theta}_{\tau}-\theta_{\tau} & =\int\widehat{\phi}_{\theta_{\tau},\eff}(V)d\mathbb{P}_{N}-\int\phi_{\theta_{\tau},\eff}(V)d\mathbb{P}\nonumber \\
 & =\int\phi_{\theta_{\tau},\eff}(V)d\mathbb{P}_{N}\label{eq:asym_part1}\\
 & +\int\{\widehat{\phi}_{\theta_{\tau},\eff}(V)-\phi_{\theta_{\tau},\eff}(V)\}d\mathbb{P}\label{eq:asym_part2}\\
 & +\int\{\widehat{\phi}_{\theta_{\tau},\eff}(V)-\phi_{\theta_{\tau},\eff}(V)\}d(\mathbb{P}_{N}-\mathbb{P}),\label{eq:asym_part3}
\end{align}
where the first term (\ref{eq:asym_part1}) is asymptotically normal
by the central limit theorem. The third term (\ref{eq:asym_part3})
is the empirical process which is negligible if $\phi_{\theta_{\tau},\eff}(V)$
belongs to Donsker classes or the cross-fitting technique is employed.
Under the assumptions in Theorem \ref{thm:theta_dr}, and the regularity conditions \ref{assump:rate}, we have 
\[
\widehat{\theta}_{\tau}=\theta_{\tau}+\frac{1}{N}\sum_{i\in\mathcal{R}\cup\mathcal{E}}\phi_{\theta_{\tau},\eff}(V)+\|{\rm Rem}(\widehat{\mathbb{P}},\mathbb{P})\|_{L_{2}}+o_{\pr}(N^{-1/2}),
\]
where $\widehat{\mathbb{P}}$ is the estimated counterpart of the
true distribution $\mathbb{P}$, and 
\[
\|{\rm Rem}(\widehat{\mathbb{P}},\mathbb{P})\|_{L_{2}}^{2}=\int\{\widehat{\phi}_{\theta_{\tau},\eff}(V)-\phi_{\theta_{\tau},\eff}(V)\}^{2}d\mathbb{P}
\]
is second-order remainder term (\ref{eq:asym_part2}). By the definition
that $\phi_{\theta_{\tau},\eff}(V)=\int_{0}^{\tau}\{\phi_{S_{1},\eff}(t,V)-\phi_{S_{0},\eff}(t,V)\}dt$,
we will first characterize the remainder term induced by $\phi_{S_{0},\eff}(V)$,
the remainder term induced by $\phi_{S_{1},\eff}(V)$ follows in similar techniques. Combining parts (\ref{eq:IPW1}), (\ref{eq:IPW2}), (\ref{eq:aug_part1}),
and (\ref{eq:aug_part2}) in $\psi_{S_{0},\eff}(V)$, we can show
\begin{align*}
 & m_{1}(t,V)+\left\{ \frac{\Delta}{S^{C}(Y\mid X,R=1)}-1\right\} m_{1}(t,V)+m_{2}(t,V)+\left\{ \frac{\Delta}{S^{C}(Y\mid X,R=0)}-1\right\} m_{2}(t,V)\\
 & +\int_{0}^{\infty}\frac{dM_{0}^{C}(r\mid X)}{S^{C}(r\mid X,R=1)}\E\left\{ m_{1}(t,V)\mid T>r,X\right\} +\int_{0}^{\infty}\frac{dM_{0}^{C}(r\mid X)}{S^{C}(r\mid X,R=0)}\E\left\{ m_{2}(t,V)\mid T>r,X\right\} \\
 & =m_{1}(t,V)+m_{2}(t,V)\\
 & -\int_{0}^{\infty}\frac{dM_{0}^{C}(r\mid X)}{S^{C}(r\mid X,R=1)}\left[m_{1}(t,V)-\E\left\{ m_{1}(t,V)\mid T>r,X\right\} \right]\\
 & -\int_{0}^{\infty}\frac{dM_{0}^{C}(r\mid X)}{S^{C}(r\mid X,R=0)}\left[m_{2}(t,V)-\E\left\{ m_{2}(t,V)\mid T>r,X\right\} \right],
\end{align*}
where 
\begin{align*}
 & m_{1}(t,V)=\frac{R(1-A)q_{R}(X)\left\{ \mathbf{1}(T>t)-S_{0}(t\mid R=1)\right\} }{\pr(R=1)D(t,X)},\\
 & m_{2}(t,V)=\frac{(1-R)q_{R}(X)r(t,X)\left\{ \mathbf{1}(T>t)-S_{0}(t\mid R=1)\right\} }{\pr(R=1)D(t,X)},\\
 & 1-\frac{\Delta}{S^{C}(Y\mid X,R=r)}=\int_{0}^{\infty}\frac{dM_{0}^{C}(r\mid X)}{S^{C}(r\mid X,R=1)},
\end{align*}
by Lemma 10.4 from \citet{tsiatis2006semiparametric,zeng2007maximum}.
Combine them with (\ref{eq:aug_part_RA}), the second-order remainder
term $\int\{\widehat{\phi}_{S_{0},\eff}(V)-\phi_{S_{0},\eff}(V)\}d\mathbb{P}$
becomes
\begin{align}
 & \mathbb{P}\left(\frac{\widehat{S}_{0}(t\mid X,R=1)-S_{0}(t\mid X,R=1)}{\pr(R=1)\widehat{D}(t,X)\{1-\widehat{\pi}_{R}(X)\}}\left[\pi_{R}(X)\{\pi_{A}(X)-\widehat{\pi}_{A}(X)\}\widehat{\pi}_{R}(X)+\widehat{r}(t,X)\{\pi_{R}(X)-\widehat{\pi}_{R}(X)\}\right]\right)\nonumber \\
 & +\mathbb{P}\left(\int_{0}^{\infty}\frac{d\widehat{M}_{0}^{C}(r\mid X)}{\widehat{S}^{C}(r\mid X,R=1)}\left[\widehat{m}_{1}(t,V)-\widehat{\E}\left\{ \widehat{m}_{1}(t,V)\mid T>r,X\right\} \right]\right)\label{eq:rem-part1}\\
 & +\mathbb{P}\left(\int_{0}^{\infty}\frac{d\widehat{M}_{0}^{C}(r\mid X)}{\widehat{S}^{C}(r\mid X,R=0)}\left[\widehat{m}_{2}(t,V)-\widehat{\E}\left\{ \widehat{m}_{2}(t,V)\mid T>r,X\right\} \right]\right),\label{eq:rem-part2}
\end{align}
since $\mathbb{P}\{m_{1}(t,V)\}=\mathbb{P}\{m_{2}(t,V)\}=0$, where
\begin{align*}
\widehat{m}_{1}(t,V) & =\frac{R(1-A)\widehat{q}_{R}(X)\left\{ \mathbf{1}(T>t)-S_{0}(t\mid R=1)\right\} }{\pr(R=1)\widehat{D}(t,X)},\\
\widehat{m}_{2}(t,V) & =\frac{(1-R)\widehat{q}_{R}(X)\widehat{r}(t,X)\left\{ \mathbf{1}(T>t)-S_{0}(t\mid R=1)\right\} }{\pr(R=1)\widehat{D}(t,X)}.
\end{align*}
 Next, we compute the expectations of $d\widehat{M}_{0}^{C}(r\mid X)$
and $\E\left\{ \mathbf{1}(T>t)\mid T>r,X\right\} $ under the true
distribution $\mathbb{P}$ by conditioning on $\{T>r,C>r,X\}$:
\begin{align*}
 & \mathbb{P}\{d\widehat{M}_{0}^{C}(r\mid X)\mid T>r,C>r,X\}=\mathbf{1}(Y>r)\{\lambda_{0}^{C}(r\mid X)-\widehat{\lambda}_{0}^{C}(r\mid X)\}dr,\\
 & \mathbb{P}[\E\{\mathbf{1}(T>t)\mid T>r,C>r,X\}]=\mathbb{P}[\E\{\mathbf{1}(T>t)\mid T>r,X\}],
\end{align*}
where the first equality holds since $\mathbb{P}\{\mathbf{1}(Y>r,\Delta=0,A=0)\mid T>r,C>r,X\}=\mathbf{1}(Y>r)\lambda_{0}^{C}(r\mid X)$,
and the second equality holds under Assumption \ref{assum:censoring_noninformative}.
By
conditioning on $\{T>r,C>r,X\}$ for $r<t$, we show that the expectation (\ref{eq:rem-part1}) is equal to 
\begin{align*}
 & \mathbb{P}\left(\int_{0}^{\infty}\frac{\mathbf{1}(Y>r)\pi_{R}(X)\{1-\pi_{A}(X)\}\widehat{q}_{R}(X)}{\pr(R=1)\widehat{D}(t,X)\widehat{S}^{C}(r\mid X,R=1)}\{\lambda_{0}^{C}(r\mid X,R=1)-\widehat{\lambda}_{0}^{C}(r\mid X,R=1)\}dr\right.\\
 & \left.\times\left[\E\left\{ \mathbf{1}(T>t)\mid T>r,X\right\} -\widehat{\E}\left\{ \mathbf{1}(T>t)\mid T>r,X\right\} \right]\right).
\end{align*}
Similar iterated expectation can be applied to (\ref{eq:rem-part2}).
Under the regularity conditions \ref{assump:rate}, we collect all the terms above and use the Cauchy-Schwarz
inequality:
\begin{align*}
 & \int|\widehat{\phi}_{S_{0},\eff}(V)-\phi_{S_{0},\eff}(V)|d\mathbb{P}\\
 & \lesssim\left\{ \|\widehat{\pi}_{A}(X)-\pi_{A}(X)\|_{L_{2}}+\|\widehat{\pi}_{R}(X)-\pi_{R}(X)\|_{L_{2}}+\max_{r<t}\|\widehat{\lambda}_{0}^{C}(r\mid X,R=1)-\lambda_{0}^{C}(r\mid X,R=1)\|_{L_{2}}\right\} \\
 & \times\|\widehat{S}_{0}(t\mid X,R=1)-S_{0}(t\mid X,R=1)\|_{L_{2}},
\end{align*}
where $\lesssim$ indicates that the inequality holds up to a multiplicative
constant. Thus, we can show the second-order remainder term is bounded:
\begin{align*}
&\|{\rm Rem}(\widehat{\mathbb{P}},\mathbb{P})\|_{L_{2}}^{2}  =\int\{\widehat{\phi}_{\theta_{\tau},\eff}(V)-\phi_{\theta_{\tau},\eff}(V)\}^{2}d\mathbb{P}\\
 & \lesssim\left\{ \|\widehat{S}_{0}(t\mid X,R=1)-S_{0}(t\mid X,R=1)\|_{L_{2}}+\|\widehat{S}_{1}(t\mid X,R=1)-S_{1}(t\mid X,R=1)\|_{L_{2}}\right\} \\
 & \times\left\{ \|\widehat{\pi}_{A}(X)-\pi_{A}(X)\|_{L_{2}}+\|\widehat{\pi}_{R}(X)-\pi_{R}(X)\|_{L_{2}}+\max_{r<t}\|\widehat{\lambda}_{0}^{C}(r\mid X,R)-\lambda_{0}^{C}(r\mid X,R)\|_{L_{2}}\right\} ,
\end{align*}
which is negligible under the regularity conditions \ref{assump:rate}. Thus, we have $\widehat{\theta}_{\tau}=\theta_{\tau}+\sum_{i\in\mathcal{R}\cup\mathcal{E}}\phi_{\theta_{\tau},\eff}(V)/N+o_{\pr}(N^{-1/2})$,
which achieves semiparametric efficiency $\mathbb{V}_{\tau}=\E\{\psi_{\theta_{\tau},\eff}^{2}(V)\}$.

\subsection{Proof of Lemma \ref{lemma:pseudo-outcomes}}\label{sec:pseudo}

Follow the similar technique in the proof of Theorem \ref{thm:EIF0},
the observed-data EIF-motivated estimator for $S_{0}(t\mid X,R=1)$
with the trial data only 
\begin{align*}
\kappa_{0}(t,V\mid R=1) & =S_{0}(t\mid X,R=1)\\
 & +\frac{R(1-A)\Delta\{\mathbf{1}(Y>t)-S_{0}(t\mid X,R=1)\}}{\pi_{R}(X)\{1-\pi_{A}(X)\}S^{C}(Y\mid X,R=1)}\\
 & +\int_{0}^{\infty}\frac{R(1-A)dM_{0}^{C}(r\mid X,R=1)}{\pi_{R}(X)\{1-\pi_{A}(X)\}S^{C}(r\mid X,R=1)}\E\{\mathbf{1}(T>t)-S_{0}(t\mid X,R=1)\mid T>r\}.
\end{align*}
By simple algebra, we obtain the simplified formula in the main paper
\begin{align*}
\kappa_{0}(t,V\mid R=1) & =S_{0}(t\mid X,R=1)\\
 & +\frac{R(1-A)\Delta\{\mathbf{1}(Y>t)-S_{0}(t\mid X)\}}{\pi_{R}(X)\{1-\pi_{A}(X)\}S^{C}(Y\mid X,R=1)}\\
 & +\int_{0}^{t}\frac{R(1-A)dM_{0}^{C}(r\mid X,R=1)}{\pi_{R}(X)\{1-\pi_{A}(X)\}S^{C}(r\mid X,R=1)}\frac{S_{0}(t\mid X,R=1)}{S_{0}(r\mid X,R=1)}\\
 & +\frac{R(1-A)\mathbf{1}(Y>t)}{\pi_{R}(X)\{1-\pi_{A}(X)\}}\left\{ \frac{1}{S^{C}(t\mid X,R=1)}-\frac{\Delta}{S^{C}(Y\mid X,R=1)}\right\} \\
 & -\frac{R(1-A)S_{0}(t\mid X)}{\pi_{R}(X)\{1-\pi_{A}(X)\}}\left\{ 1-\frac{\Delta}{S^{C}(Y\mid X,R=1)}\right\} \\
 & =S_{0}(t\mid X,R=1)\\
 & +\frac{R(1-A)}{\pi_{R}(X)\{1-\pi_{A}(X)\}}\left\{ \frac{\mathbf{1}(Y>t)}{S^{C}(t\mid X_{i},R=1)}-S_{0}(t\mid X,R=1)\right\} \\
 & +\int_{0}^{t}\frac{R(1-A)dM_{0}^{C}(r\mid X,R=1)}{\pi_{R}(X)\{1-\pi_{A}(X)\}S^{C}(r\mid X,R=1)}\frac{S_{0}(t\mid X,R=1)}{S_{0}(r\mid X,R=1)},
\end{align*}
where 
\begin{align*}
\int_{t}^{\infty}\frac{dM_{0}^{C}(r\mid X,R=1)}{S^{C}(r\mid X,R=1)} & =\mathbf{1}(Y>t)\left\{ \frac{1}{S^{C}(t\mid X,R=1)}-\frac{\Delta}{S^{C}(Y\mid X,R=1)}\right\} ,\\
\int_{0}^{\infty}\frac{dM_{0}^{C}(r\mid X,R=1)}{S^{C}(r\mid X,R=1)} & =1-\frac{\Delta}{S^{C}(Y\mid X,R=1)},
\end{align*}
by our arguments in Theorem \ref{thm:theta_dr}. Following another
representation of $\kappa_{0}(t,V\mid R=1)$, we can show that 
\begin{align*}
\kappa_{0}(t,V\mid R=1) & =S_{0}(t\mid X,R=1)\\
 & +\frac{R(1-A)\{\mathbf{1}(T>t)-S_{0}(t\mid X)\}}{\pi_{R}(X)\{1-\pi_{A}(X)\}}\\
 & +\left\{ \frac{\Delta}{S^{C}(Y\mid X,R=1)}-1\right\} \frac{R(1-A)\{\mathbf{1}(T>t)-S_{0}(t\mid X,R=1)\}}{\pi_{R}(X)\{1-\pi_{A}(X)\}}\\
 & +\int_{0}^{\infty}\frac{R(1-A)dM_{0}^{C}(r\mid X,R)}{\pi_{R}(X)\{1-\pi_{A}(X)\}S^{C}(r\mid X,R=1)}\E\{\mathbf{1}(T>t)-S_{0}(t\mid X)\mid T>r,R=1\}\\
 & =S_{0}(t\mid X,R=1)+\frac{R(1-A)\{\mathbf{1}(T>t)-S_{0}(t\mid X,R=1)\}}{\pi_{R}(X)\{1-\pi_{A}(X)\}}\\
 & +\int_{0}^{\infty}\frac{R(1-A)dM_{0}^{C}(r\mid X,R)}{\pi_{R}(X)\{1-\pi_{A}(X)\}S^{C}(r\mid X,R=1)}\left[\E\{\mathbf{1}(T>t)\mid T>r,R=1\}-\mathbf{1}(T>t)\right],
\end{align*}
where $1-\Delta/S^{C}(Y\mid X,R=r)=\int_{0}^{\infty}dM_{0}^{C}(r\mid X)/S^{C}(r\mid X,R=1)$.
Let $\kappa_{0}^{*}(t,V\mid R=1)$ be the probability limit of $\widehat{\kappa}_{0}(t,V\mid R=1)$,
we can show that
\begin{align*}
 & \E\{\kappa_{0}^{*}(t,V\mid R=1)-S_{0}(t\mid X,R=1)\mid X\}\\
\leq & \left\{ S_{0}^{*}(t\mid X,R=1)-S_{0}(t\mid X,R=1)\right\} \cdot\left[1-\frac{\pi_{R}(X)\{1-\pi_{A}(X)\}}{\pi_{R}^{*}(X)\{1-\pi_{A}^{*}(X)\}}\right]\\
 & +\frac{\pi_{R}(X)\{1-\pi_{A}(X)\}}{\pi_{R}^{*}(X)\{1-\pi_{A}^{*}(X)\}}\E\left(\int_{0}^{\infty}\frac{\E\{dM_{0}^{C*}(r\mid X,R=1)\mid C>r,T>r\}}{S^{C*}(r\mid X,R=1)}\right.\\
 & \left.\times\left[\E^{*}\{\mathbf{1}(T>t)\mid T>r\}-\E\{\mathbf{1}(T>t)\mid T>r\}\right]\right).
\end{align*}
Thus, we establish the bound for $\kappa_{0}^{*}(t,V\mid R=1)$ by
\begin{align*}
 & \|\E\{\kappa_{0}^{*}(t,V\mid R=1)-S_{0}(t\mid X,R=1)\mid X\}\|_{L_{2}}\\
 & \lesssim\|\pi_{R}^{*}(X)-\pi_{R}(X)\|_{L_{2}}\cdot\|S_{0}^{*}(t\mid X,R=1)-S_{0}(t\mid X,R=1)\|_{L_{2}}\\
 & +\|\pi_{A}^{*}(X)-\pi_{A}(X)\|_{L_{2}}\cdot\|S_{0}^{*}(t\mid X,R=1)-S_{0}(t\mid X,R=1)\|_{L_{2}}\\
 & +\|\lambda_{0}^{C*}(t\mid X,R=1)-\lambda_{0}^{C}(t\mid X,R=1)\|\cdot\|S_{0}^{*}(t\mid X,R=1)-S_{0}(t\mid X,R=1)\|_{L_{2}}.
\end{align*}
Similarly, we can establish the bound for $\kappa_{0}^{*}(t,V\mid R=0)$
as 
\begin{align*}
 & \|\E\{\kappa_{0}^{*}(t,V\mid R=0)-S_{0}(t\mid X,R=0)\mid X\}\|_{L_{2}}\\
 & \lesssim\|\pi_{R}^{*}(X)-\pi_{R}(X)\|_{L_{2}}\cdot\|S_{0}^{*}(t\mid X,R=0)-S_{0}(t\mid X,R=0)\|_{L_{2}}\\
 & +\|\lambda_{0}^{C*}(t\mid X,R=0)-\lambda_{0}^{C}(t\mid X,R=0)\|_{L_{2}}\cdot\|S_{0}^{*}(t\mid X,R=0)-S_{0}(t\mid X,R=0)\|_{L_{2}}.
\end{align*}
Putting the bounds for $\kappa_{0}^{*}(t,V\mid R=1)$ and $\kappa_{0}^{*}(t,V\mid R=0)$
together, we obtain the desired result: 
\begin{align*}
 & \|\E\{\xi(V)-b_{0}\mid X\}\|_{L_{2}}\\
 & \lesssim\sum_{r=0}^{1}\|\pi_{R}^{*}(X)-\pi_{R}(X)\|_{L_{2}}\cdot\|S_{0}^{*}(t\mid X,R=r)-S_{0}(t\mid X,R=r)\|_{L_{2}}\\
 & +\sum_{r=0}^{1}\|S^{C*}(t\mid X,R=r)-S^{C}(t\mid X,R=r)\|_{L_{2}}\cdot\|S_{0}^{*}(t\mid X,R=r)-S_{0}(t\mid X,R=r)\|_{L_{2}}\\
 & +\|\pi_{A}^{*}(X)-\pi_{A}(X)\|_{L_{2}}\cdot\|S_{0}^{*}(t\mid X,R=1)-S_{0}(t\mid X,R=1)\|_{L_{2}},
\end{align*}
which completes the proof of Lemma \ref{lemma:pseudo-outcomes}. 

\subsection{Proof of Theorem \ref{thm:eff_gain}}\label{sec:proof_gain}

Under the condition that there exists a subset $\mathcal{A}$of the
external controls such that $S(t\mid X_{i},R=1)=S(t\mid X_{i},R=0)$
for any time $t$ and subject $i\in\mathcal{A}$, the tangent space
$H_{4}$ in (\ref{eq:tangent_space}) is modified for the updated
restricted moment condition
\begin{align*}
H_{4}^{*} & =\left\{ \Gamma(T,X,R,A):\E\{\Gamma(T,X,R,A)\mid X,R,A\}=0\right\} \\
 & \cap\left\{ \Gamma(Y,X,R,A):\E\left[\left\{ \frac{(1-R)\mathbf{1}(b=0)\mathbf{1}(T>t)}{\pr(R=0,b=0\mid X)}-\frac{R(1-A)\mathbf{1}(T>t)}{\pr(R=1,A=0\mid X)}\right\} \Gamma(Y,X,R,A)\mid X\right]=0,t<\tau\right\} .
\end{align*}
Similarly, we find the proper functions $C_{1}^{*}$ and $C_{2}^{*}$
as in (\ref{eq:C1_C2}) to obtain the full-data EIF for $S_{0}(t\mid R=1)$:
\begin{align*}
\psi_{S_{0},\eff}^{F,\mathcal{A}}(t,W) & =\frac{R(1-A)}{\pr(R=1)}\frac{q_{R}(X)\left\{ \mathbf{1}(T>t)-S_{0}(t\mid X,R=1)\right\} }{D_{b_{0}}(t,X)}\\
 & +\frac{(1-R)\mathbf{1}(b_{0}=0)}{\pr(R=1)}\frac{r(t,X)q_{R}(X)\left\{ \mathbf{1}(T>t)-S_{0}(t\mid X,R=1)\right\} }{D_{b_{0}}(t,X)}\\
 & +\frac{R}{\pr(R=1)}\{S_{0}(t\mid X,R=1)-S_{0}(t\mid R=1)\},\\
 & =\frac{R(1-A)}{\pr(R=1)}\frac{q_{R}(X)\left\{ \mathbf{1}(T>t)-S_{0}(t\mid R=1)\right\} }{D_{b_{0}}(t,X)}\\
 & +\frac{(1-R)\mathbf{1}(b_{0}=0)q_{R}(X)}{\pr(R=1)}\frac{r(t,X)\left\{ \mathbf{1}(T>t)-S_{0}(t\mid R=1)\right\} }{D_{b_{0}}(t,X)}\\
 & +\frac{R\{A-\pi_{A}(X)\}q_{R}(X)}{\pr(R=1)D_{b_{0}}(t,X)}\{S_{0}(t\mid X,R=1)-S_{0}(t\mid R=1)\}\\
 & +\frac{r(t,X)\{R\pr(b=0\mid X,R=0)-(1-R)\mathbf{1}(b_{0}=0)q_{R}(X)\}}{\pr(R=1)D_{b_{0}}(t,X)}\{S_{0}(t\mid X,R=1)-S_{0}(t\mid R=1)\},
\end{align*}
where $D_{b_{0}}(t,X)=r(t,X)P(b=0\mid X,R=0)+\{1-\pi_{A}(X)\}q_{R}(X)$.
By finding the optimal element of the augmentation space with some
algebra similar to Section \ref{subsec:observed-EIF}, we obtain the
modified observed-data EIF under the updated restricted moment condition
\begin{align*}
\psi_{S_{0},\eff}^{\mathcal{A}}(t,V) & =\frac{R(1-A)}{\pr(R=1)}\frac{q_{R}(X)\mathbf{1}(Y>t)}{S^{C}(t\mid X,R=1)D_{b_{0}}(t,X)}\\
 & +\frac{(1-R)\mathbf{1}(b_{0}=0)}{\pr(R=1)}\frac{q_{R}(X)r(t,X)\mathbf{1}(Y>t)}{S^{C}(t\mid X,R=0)D_{b_{0}}(t,X)}\\
 & +\int_{0}^{t}\frac{R(1-A)}{\pr(R=1)}\frac{q_{R}(X)dM_{0}^{C}(r\mid X)}{S^{C}(r\mid X,R=1)D_{b_{0}}(t,X)}\frac{S_{0}(t\mid X)}{S_{0}(r\mid X)}\\
 & +\int_{0}^{t}\frac{(1-R)\mathbf{1}(b_{0}=0)}{\pr(R=1)}\frac{q_{R}(X)r(t,X)dM_{0}^{C}(r\mid X)}{S^{C}(r\mid X,R=0)D_{b_{0}}(t,X)}\frac{S_{0}(t\mid X)}{S_{0}(r\mid X)}\\
 & +\frac{Rq_{R}(X)\{A-\pi_{A}(X)\}S_{0}(t\mid X)}{\pr(R=1)D_{b_{0}}(t,X)}\\
 & +\frac{r(t,X)\{R\pr(b=0\mid X,R=0)-(1-R)\mathbf{1}(i\in\mathcal{A})q_{R}(X)\}S_{0}(t\mid X)}{\pr(R=1)D_{b_{0}}(t,X)}-\frac{RS_{0}(t\mid R=1)}{\pr(R=1)}\\
 & =\phi_{S_{0},\eff}^{\mathcal{A}}(t,V)-\frac{RS_{0}(t\mid R=1)}{\pr(R=1)},
\end{align*}
which belongs to the orthogonal complement of the updated observed-data
nuisance tangent space $\Lambda_{\eta}^{*\perp}$. From our previous arguments, we know that the trial-only efficient influence function
is $\psi_{S_{0},\eff}^{\text{rct}}(t,V)$:
\begin{align*}
\psi_{S_{0},\eff}^{\text{rct}}(t,V) & =\frac{R(1-A)\mathbf{1}(Y>t)}{\pr(R=1)\{1-\pi_{A}(X)\}S^{C}(t\mid X,R=1)}\\
 & +\int_{0}^{t}\frac{R(1-A)dM_{0}^{C}(r\mid X)}{\pr(R=1)\{1-\pi_{A}(X)\}S^{C}(r\mid X,R=1)}\frac{S_{0}(t\mid X)}{S_{0}(r\mid X)}\\
 & +\frac{R\{A-\pi_{A}(X)\}S_{0}(t\mid X,R=1)}{\pr(R=1)\{1-\pi_{A}(X)\}}-\frac{RS_{0}(t\mid R=1)}{\pr(R=1)},
\end{align*}
which leads to the EIF of $\theta_{\tau}$ by $\psi_{\theta_{\tau},\eff}^{\text{rct}}(V)=\int_{0}^{\tau}\{\psi_{S_{1},\eff}(t,V)-\psi_{S_{0},\eff}^{\text{rct}}(t,V)\}dt$
with the asymptotic variance $\mathbb{V}_{\tau}^{\text{aipw}}$. Compare
the asymptotic variance $\mathbb{V}_{\tau}^{\text{adapt}}$ to $\mathbb{V}_{\tau}^{\text{aipw}}$,
we have 
\begin{align}
\mathbb{V}_{\tau}^{\text{aipw}}-\mathbb{V}_{\tau}^{\text{adapt}} & =\E\{\psi_{\theta_{\tau},\eff}^{\text{rct}}(V)\}^{2}-\E\{\psi_{\theta_{\tau},\eff}^{\mathcal{A}}(V)\}^{2}\nonumber \\
 & =\E\{\psi_{\theta_{\tau},\eff}^{\text{rct}}(V)-\psi_{\theta_{\tau},\eff}^{\mathcal{A}}(V)\}^{2}\nonumber \\
 & +2\E[\psi_{\theta_{\tau},\eff}^{\mathcal{A}}(t,V)\{\psi_{\theta_{\tau},\eff}^{\text{rct}}(V)-\psi_{\theta_{\tau},\eff}^{\mathcal{A}}(V)\}].\label{eq:eff_var_part2}
\end{align}
By directional derivative, we have
\[
\dot{\theta}_{\tau}=\E\{\psi_{\theta_{\tau},\eff}^{\text{rct}}(V)s(V)\}=\E\{\psi_{\theta_{\tau},\eff}^{\mathcal{A}}(V)s(V)\},
\]
where $s(V)$ is the score function for the observed data, and therefore
$\E[\{\psi_{\theta_{\tau},\eff}^{\text{rct}}(V)-\psi_{\theta_{\tau},\eff}^{\mathcal{A}}(V)\}s(V)]=0$,
implying $\psi_{\theta_{\tau},\eff}^{\text{rct}}(V)-\psi_{\theta_{\tau},\eff}^{\mathcal{A}}(V)$
belongs to the updated observed-data nuisance tangent space $\Lambda_{\eta}^{*}$.
Note that $\psi_{\theta_{\tau},\eff}^{\mathcal{A}}(t,V)\in\Lambda_{\eta}^{*\perp}$,
we have (\ref{eq:eff_var_part2}) equals to zero, and 
\[
\mathbb{V}_{\tau}^{\text{aipw}}-\mathbb{V}_{\tau}^{\text{adapt}}=\E\{\psi_{\theta_{\tau},\eff}^{\text{rct}}(V)-\psi_{\theta_{\tau},\eff}^{\mathcal{A}}(V)\}^{2},
\]
where $\psi_{\theta_{\tau},\eff}^{\text{rct}}-\psi_{\theta_{\tau},\eff}^{\mathcal{A}}(V)=\int_{0}^{\tau}\psi_{S_{0},\eff}^{\text{rct}}(t,V)dt-\int_{0}^{\tau}\psi_{S_{0},\eff}^{\mathcal{A}}(t,V)dt$.
Next, we can show 
\begin{align*}
 & \psi_{S_{0},\eff}^{\text{rct}}(t,V)-\psi_{S_{0},\eff}^{\mathcal{A}}(t,V)\\
 & =\frac{R(1-A)}{\pr(R=1)S^{C}(t\mid X,R=1)}\left\{ \frac{1}{1-\pi_{A}(X)}-\frac{q_{R}(X)}{D_{b_{0}}(t,X)}\right\} \mathbf{1}(Y>t)\\
 & -\int_{0}^{t}\frac{R(1-A)dM_{0}^{C}(r\mid X)}{\pr(R=1)S^{C}(t\mid X,R=1)}\frac{S_{0}(t\mid X)}{S_{0}(r\mid X)}\left\{ \frac{q_{R}(X)}{D_{b_{0}}(t,X)}-\frac{1}{1-\pi_{A}(X)}\right\} \\
 & -\frac{(1-R)\mathbf{1}(b_{0}=0)}{\pr(R=1)}\frac{q_{R}(X)r(t,X)\mathbf{1}(Y>t)}{S^{C}(t\mid X,R=0)D_{b_{0}}(t,X)}\\
 & -\int_{0}^{t}\frac{(1-R)\mathbf{1}(b_{0}=0)}{\pr(R=1)}\frac{q_{R}(X)r(t,X)dM_{0}^{C}(r\mid X)}{S^{C}(r\mid X,R=0)D_{b_{0}}(t,X)}\frac{S_{0}(t\mid X)}{S_{0}(r\mid X)}\\
 & +\frac{R\{A-\pi_{A}(X)\}S_{0}(t\mid X,R=1)}{\pr(R=1)\{1-\pi_{A}(X)\}}-\frac{Rq_{R}(X)\{A-\pi_{A}(X)\}S_{0}(t\mid X,R=1)}{\pr(R=1)D_{b_{0}}(t,X)}\\
 & -\frac{r(t,X)\{R\pr(b=0\mid X,R=0)-(1-R)\mathbf{1}(b_{0}=0)q_{R}(X)\}S_{0}(t\mid X,R=1)}{\pr(R=1)D_{b_{0}}(t,X)}.
\end{align*}
By some algebra, we have 
\begin{align*}
 & \frac{R\{A-\pi_{A}(X)\}S_{0}(t\mid X,R=1)}{\pr(R=1)\{1-\pi_{A}(X)\}}=\frac{R}{\pr(R=1)}\left\{ 1-\frac{1-A}{1-\pi_{A}(X)}\right\} S_{0}(t\mid X,R=1),
\end{align*}
and 
\begin{align*}
 & \frac{Rq_{R}(X)\{A-\pi_{A}(X)\}S_{0}(t\mid X,R=1)}{\pr(R=1)D_{b_{0}}(t,X)}\\
 & +\frac{r(t,X)\{R\pr(b=0\mid X,R=0)-(1-R)\mathbf{1}(b_{0}=0)q_{R}(X)\}S_{0}(t\mid X,R=1)}{\pr(R=1)D_{b_{0}}(t,X)}\\
 & \frac{1}{\pr(R=1)}\left\{ R-\frac{R(1-A)q_{R}(X)}{\pr(R=1)D_{b_{0}}(t,X)}-\frac{r(t,X)(1-R)\mathbf{1}(b_{0}=0)q_{R}(X)}{\pr(R=1)D_{b_{0}}(t,X)}\right\} S_{0}(t\mid X,R=1).
\end{align*}
Plugging these terms back to $\psi_{S_{0},\eff}^{\text{rct}}(t,V)-\psi_{S_{0},\eff}^{\mathcal{A}}(t,V)$,
we have

\begin{align}
 & \psi_{S_{0},\eff}^{\text{rct}}(t,V)-\psi_{S_{0},\eff}^{\mathcal{A}}(t,V)\nonumber \\
 & =\frac{R(1-A)}{\pr(R=1)S^{C}(t\mid X,R=1)}\left\{ \frac{1}{1-\pi_{A}(X)}-\frac{q_{R}(X)}{D_{b_{0}}(t,X)}\right\} \mathbf{1}(Y>t)\label{eq:eff_proof_part1}\\
 & -\int_{0}^{t}\frac{R(1-A)dM_{0}^{C}(r\mid X)}{\pr(R=1)S^{C}(t\mid X,R=1)}\frac{S_{0}(t\mid X)}{S_{0}(r\mid X)}\left\{ \frac{q_{R}(X)}{D_{b_{0}}(t,X)}-\frac{1}{1-\pi_{A}(X)}\right\} \label{eq:eff_proof_part2}\\
 & -\frac{(1-R)\mathbf{1}(b_{0}=0)}{\pr(R=1)}\frac{q_{R}(X)r(t,X)\mathbf{1}(Y>t)}{S^{C}(t\mid X,R=0)D_{b_{0}}(t,X)}\label{eq:eff_proof_part3}\\
 & -\int_{0}^{t}\frac{(1-R)\mathbf{1}(b_{0}=0)}{\pr(R=1)}\frac{q_{R}(X)r(t,X)dM_{0}^{C}(r\mid X)}{S^{C}(r\mid X,R=0)D_{b_{0}}(t,X)}\frac{S_{0}(t\mid X)}{S_{0}(r\mid X)}\label{eq:eff_proof_part4}\\
 & -\frac{R(1-A)}{\pr(R=1)}\left\{ \frac{1}{1-\pi_{A}(X)}-\frac{q_{R}(X)}{D_{b_{0}}(t,X)}\right\} S_{0}(t\mid X)\label{eq:eff_proof_part5}\\
 & +\frac{r(t,X)(1-R)\mathbf{1}(b_{0}=0)q_{R}(X)}{\pr(R=1)D_{b_{0}}(t,X)}S_{0}(t\mid X).\label{eq:eff_proof_part6}
\end{align}
Combining (\ref{eq:eff_proof_part1}), (\ref{eq:eff_proof_part2}),
and (\ref{eq:eff_proof_part5}) gives us
\begin{align*}
 & \frac{R(1-A)}{\pr(R=1)S^{C}(t\mid X,R=1)}\left\{ \frac{1}{1-\pi_{A}(X)}-\frac{q_{R}(X)}{D_{b_{0}}(t,X)}\right\} \mathbf{1}(Y>t)\\
 & -\int_{0}^{t}\frac{R(1-A)dM_{0}^{C}(r\mid X)}{\pr(R=1)S^{C}(t\mid X,R=1)}\frac{S_{0}(t\mid X)}{S_{0}(r\mid X)}\left\{ \frac{q_{R}(X)}{D_{b_{0}}(t,X)}-\frac{1}{1-\pi_{A}(X)}\right\} \\
 & -\frac{R(1-A)}{\pr(R=1)}\left\{ \frac{1}{1-\pi_{A}(X)}-\frac{q_{R}(X)}{D_{b_{0}}(t,X)}\right\} S_{0}(t\mid X)\\
= & \frac{R(1-A)}{\pr(R=1)}\left\{ \frac{r(t,X)P(b=0\mid X,R=0)}{\{1-\pi_{A}(X)\}D_{b_{0}}(t,X)}\right\} m_{1}^{*}(t,V),
\end{align*}
where 
\begin{align*}
m_{1}^{*}(t,V) & =\frac{\mathbf{1}(Y>t)}{S^{C}(t\mid X,R=1)}+\int_{0}^{\tau}\frac{dM_{0}^{C}(r\mid X)}{S^{C}(r\mid X,R=1)}\frac{S_{0}(t\mid X)}{S_{0}(r\mid X)}-S_{0}(t\mid X)\\
 & =\frac{\Delta\mathbf{1}(Y>t)}{S^{C}(Y\mid X,R=1)}+\int_{0}^{\infty}\frac{dM_{0}^{C}(r\mid X)}{S^{C}(r\mid X,R=1)}\frac{S_{0}(t\mid X)}{S_{0}(r\mid X)}-S_{0}(t\mid X).
\end{align*}
Similarly, (\ref{eq:eff_proof_part3}), (\ref{eq:eff_proof_part4}),
and (\ref{eq:eff_proof_part6}) together gives us

\begin{align*}
 & \frac{(1-R)\mathbf{1}(b_{0}=0)}{\pr(R=1)}\frac{q_{R}(X)r(t,X)\mathbf{1}(Y>t)}{S^{C}(t\mid X,R=0)D_{b_{0}}(t,X)}\\
 & +\int_{0}^{t}\frac{(1-R)\mathbf{1}(b_{0}=0)}{\pr(R=1)}\frac{q_{R}(X)r(t,X)dM_{0}^{C}(r\mid X)}{S^{C}(r\mid X,R=0)D_{b_{0}}(t,X)}\frac{S_{0}(t\mid X)}{S_{0}(r\mid X)}\\
 & -\frac{r(t,X)(1-R)\mathbf{1}(b_{0}=0)q_{R}(X)}{\pr(R=1)D_{b_{0}}(t,X)}S_{0}(t\mid X)\\
= & \frac{(1-R)\mathbf{1}(b_{0}=0)}{\pr(R=1)}\frac{q_{R}(X)r(t,X)}{D_{b_{0}}(t,X)}m_{2}^{*}(t,V),
\end{align*}
where 
\begin{align*}
m_{2}^{*}(t,V) & =\frac{\mathbf{1}(Y>t)}{S^{C}(t\mid X,R=0)}+\int_{0}^{t}\frac{dM_{0}^{C}(r\mid X)}{S^{C}(r\mid X,R=0)}\frac{S_{0}(t\mid X)}{S_{0}(r\mid X)}-S_{0}(t\mid X)\\
 & =\frac{\Delta\mathbf{1}(Y>t)}{S^{C}(Y\mid X,R=0)}+\int_{0}^{\infty}\frac{dM_{0}^{C}(r\mid X)}{S^{C}(r\mid X,R=0)}\frac{S_{0}(t\mid X)}{S_{0}(r\mid X)}-S_{0}(t\mid X).
\end{align*}
Then, we can show that $\E[\{\psi_{S_{0},\eff}^{\text{rct}}(t,V)-\psi_{S_{0},\eff}^{\mathcal{A}}(t,V)\}^{2}\mid X]$
equals to
\begin{align*}
 & \frac{\pi_{R}(X)\{1-\pi_{A}(X)\}}{\pr(R=1)^{2}}\left\{ \frac{r(t,X)\pr(b_{0}=0\mid X,R=0)}{\{1-\pi_{A}(X)\}D^{*}(t,X)}\right\} ^{2}\text{var}\left\{ m_{1}^{*}(t,V)\mid R=1,A=0\right\} \\
 & +\frac{\{1-\pi_{R}(X)\}\pr(b_{0}=0\mid X,R=0)}{\pr(R=1)^{2}}\left\{ \frac{q_{R}(X)r(t,X)}{D^{*}(t,X)}\right\} ^{2}\text{var}\left\{ m_{2}^{*}(t,V)\mid R=0,b_{0}=0\right\} \\
 & =\frac{\pi_{R}(X)r(t,X)\pr(b_{0}=0\mid X,R=0)}{\pr(R=1)^{2}D_{b_{0}}(t,X)\{1-\pi_{A}(X)\}}\frac{D_{b_{0}}^{*}(t,X)}{D_{b_{0}}(t,X)}\frac{r(t,X)}{r^{*}(t,X)}V_{R1,A0}^{*},
\end{align*}
where 
\begin{align*}
r^{*}(t,X) & =\frac{V_{R1,A0}^{*}}{V_{R0}^{*}},\quad D_{b_{0}}^{*}(t,X)=r^{*}(t,X)\pr(b=0\mid X,R=0)+\{1-\pi_{A}(X)\}q_{R}(X),\\
V_{R1,A0}^{*}= & \text{var}\left\{ m_{1}^{*}(t,V)\mid R=1,A=0\right\} ,\quad V_{R0}^{*}=\text{var}\left\{ m_{2}^{*}(t,V)\mid R=0,b_{0}=0\right\} .
\end{align*}
Thus, the proof of Theorem \ref{thm:eff_gain} is completed.

\section{Additional Simulations}\label{sec:additional}
\begin{figure}[!htbp]
\centering
\includegraphics[width=.6\linewidth]{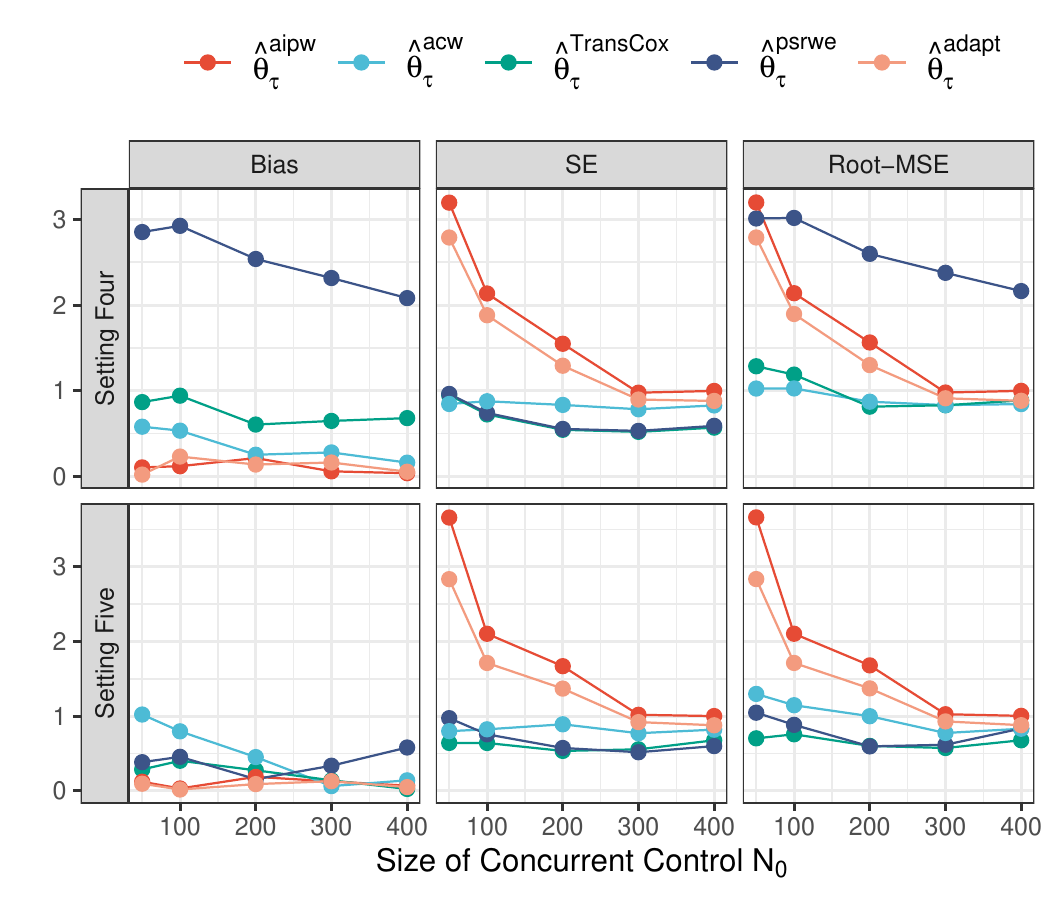}
\includegraphics[width=.35\linewidth]{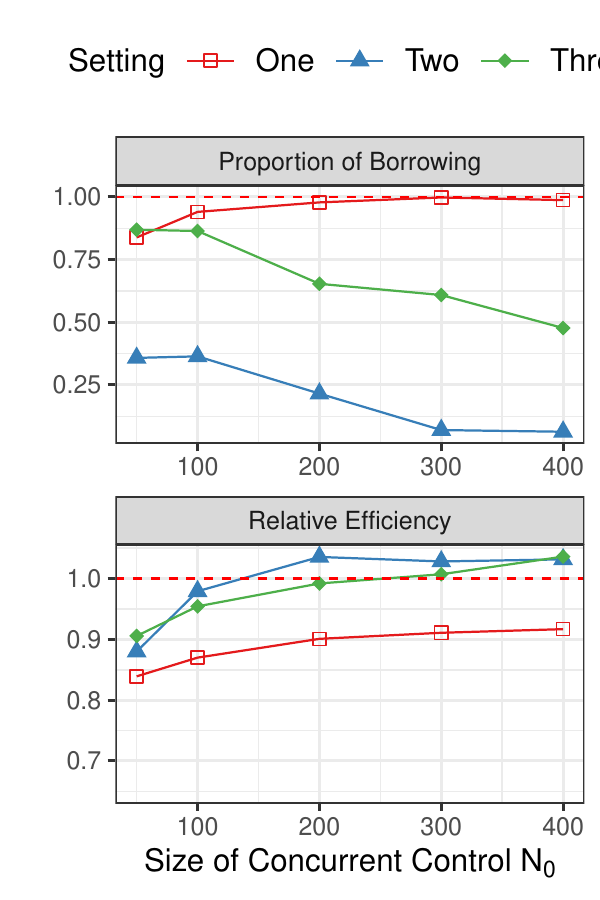}
\caption{\label{fig:sim-point-appendix} (Left) Point estimation results for RMST over $500$
Monte Carlo experiments under Settings 4) different covariate effects, and 5) different baseline time-varying hazards; (Right) Average borrowing proportion of external controls
and relative efficiency of the selective integrative estimator $\widehat{\theta}_{\tau}^{\text{adapt}}$
over $500$ Monte Carlo experiments.}
\end{figure}

\paragraph{Additional Bias-generating Settings}  
Figure \ref{fig:sim-point-appendix}(Left) presents the simulation results under Settings Four and Five. Both $\widehat{\theta}_{\tau}^{\text{adapt}}$ and $\widehat{\theta}_{\tau}^{\text{TransCox}}$ account for heterogeneity in covariate effects and the risk associated with varying baseline times in these settings. However, $\widehat{\theta}_{\tau}^{\text{TransCox}}$ is only valid under the Cox model. For example, when the conditional survival curve $S_a(t \mid X)$ does not follow the Cox model, as in Settings Two and Three of the main paper where the marginalized curves $S_a(t \mid X)$ over $U$ (or $\delta$) result in a model that no longer satisfies the Cox proportional hazards assumption, $\widehat{\theta}_{\tau}^{\text{TransCox}}$ may exhibit substantial bias, whereas the proposed estimator $\widehat{\theta}_{\tau}^{\text{adapt}}$ continues to control for bias due to its double robustness and demonstrates improved performance.

\paragraph{Asymptotic Properties of the Proposed Estimator} Figure \ref{fig:sim-point-appendix}(Right) provides more details of our proposed
selective integrative estimator, specifically focusing on its average
borrowing proportion of the external controls and its relative efficiency.
The relative efficiency is measured by the ratio of the width of its
confidence intervals to the trial-only estimator. In Setting One,
it is reasonable to observe that the selective integrative estimator
$\widehat{\theta}_{\tau}^{\text{adapt}}$ is always more efficient
compared to the benchmark as every external control is comparable,
and the proportion of borrowing approaches $1$ as $N_{0}$ increases.
Under Setting Two, the proportion of borrowing diminishes to zero
as it detects that nearly all the external controls are not comparable
when more concurrent controls become available. Under Setting Three,
the borrowing proportion approaches $50\%$, aligning well with the
true proportion of comparable external subset in our data-generation process. One side note is that our proposal might be subject to slight
relative efficiency inferiority compared to the benchmark in some
cases, which is also observed in other literature \citep{chen2021minimax}.

\begin{figure}[!htbp]
\centering
\includegraphics[width=0.49\textwidth]{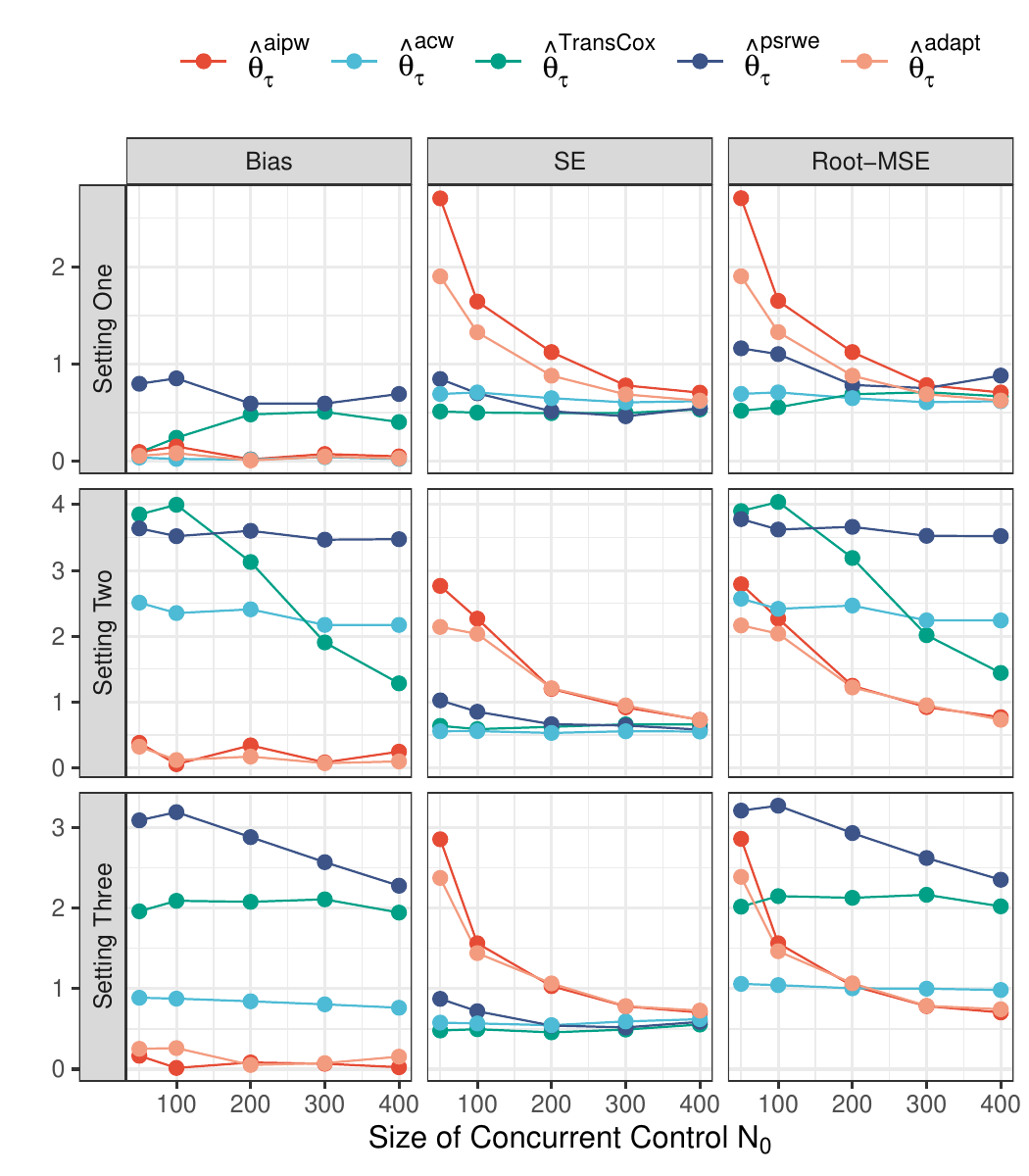}
\includegraphics[width=0.49\textwidth]{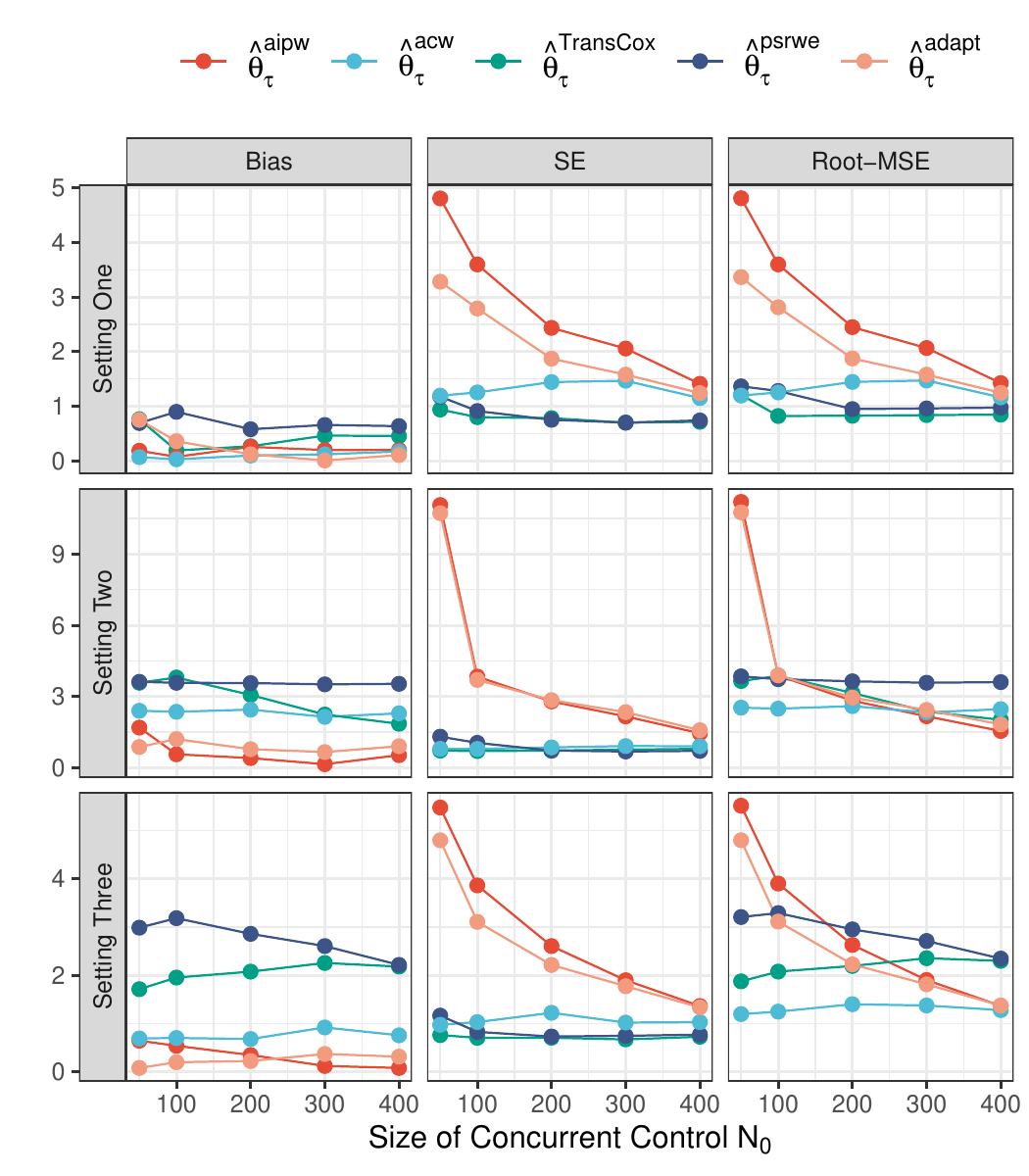}

\caption{\label{fig:sim-point-lowhigh} Point estimation results for RMST over
$500$ Monte Carlo experiments under Settings One, Two and Three when
(Left) $\beta_{C}=-2$ (low censoring rate) and (Right) $\beta_{C}=0$ (high censoring
rate).}
\end{figure}

\paragraph{Varying Censoring Intensity} We conduct additional simulation studies to evaluate our selective
borrowing estimator $\widehat{\theta}_{\tau}^{\text{adapt}}$ under
varying censoring rates for the trial. In particular, we vary the
value of $\beta_{C}$ in the hazard functions $\lambda^{C}(t\mid X,R=1)$
to represent different censoring levels, with $\beta_{C}=0$ indicating
high censoring (the censoring rate is around $60\%$) and $\beta_{C}=-2$ indicating low censoring (the censoring rate is around $20\%$). The results,
presented in Figure \ref{fig:sim-point-lowhigh}, demonstrate that
our proposed estimator effectively controls external biases across
all settings and achieves improved estimation, as shown by smaller
Root-MSEs compared to the trial-only estimator.

\end{document}